\documentclass[12pt,letterpaper]{article}
\usepackage{epsfig,rotating,setspace,latexsym,amsmath,epsf,amssymb,amsfonts,bm,theorem,cite,enumerate,longtable,accents,float,physics}
\usepackage{algorithm,algorithmic,graphicx,epsf,authblk,epstopdf,url,xcolor, soul,multirow,bbm,hyperref}
\usepackage{mathtools,comment}
\usepackage[center]{qtree}
\usepackage{tree-dvips}
\usepackage[linguistics]{forest }

\newtheorem{theorem}{Theorem}

\newtheorem{definition}{Definition}
\newtheorem{remark}{Remark}
\newtheorem{lemma}{Lemma}

\newenvironment{Proof}[1]{\medskip\par\noindent{\bf Proof:\,}\,#1}{{\mbox{\,$\blacksquare$}\par}}

\DeclareMathOperator{\diag}{diag}

\allowdisplaybreaks
\date{}

\title{Byzantine-Eavesdropper Alliance: How to Achieve Symmetric Privacy in Quantum $X$-Secure $B$-Byzantine $E$-Eavesdropped $U$-Unresponsive $T$-Colluding PIR?}

\author{Mohamed Nomeir \qquad Alptug Aytekin \qquad  Sennur Ulukus\\
	\normalsize Department of Electrical and Computer Engineering\\
	\normalsize University of Maryland, College Park, MD 20742 \\
	\normalsize \emph{mnomeir@umd.edu} \qquad      
        \emph{aaytekin@umd.edu} \qquad \emph{ulukus@umd.edu}}
 
\begin{document}

\maketitle

\vspace*{-1.0cm}

\begin{abstract}
   We consider the quantum \emph{symmetric} private information retrieval (QSPIR) problem in a system with $N$ databases and $K$ messages, with $U$ unresponsive servers, $T$-colluding servers, and $X$-security parameter, under several fundamental threat models. In the first model, there are $\mathcal{E}_1$ eavesdropped links in the uplink direction (the direction from the user to the $N$ servers), $\mathcal{E}_2$ eavesdropped links in the downlink direction (the direction from the servers to the user), where $|\mathcal{E}_1|, |\mathcal{E}_2| \leq E$; we coin this eavesdropper setting as \emph{dynamic} eavesdroppers. We show that super-dense coding gain can be achieved for some regimes. In the second model, we consider the case with Byzantine servers, i.e., servers that can coordinate to devise a plan to harm the privacy and security of the system together with static eavesdroppers, by listening to the same links in both uplink and downlink directions. It is important to note the considerable difference between the two threat models, since the eavesdroppers can take huge advantage of the presence of the Byzantine servers. Unlike the previous works in SPIR with Byzantine servers, that assume that the Byzantine servers can send only random symbols independent of the stored messages, we follow the definition of Byzantine servers in \cite{byzantine_tpir}, where the Byzantine servers can send symbols that can be functions of the storage, queries, as well as the random symbols in a way that can produce worse harm to the system. In the third and the most novel threat model, we consider the presence of Byzantine servers and dynamic eavesdroppers together. We show that having dynamic eavesdroppers along with Byzantine servers in the same system model creates more threats to the system than having static eavesdroppers with Byzantine servers.
   
   This is the first work that considers the quantum version of unresponsive and eavesdropped threat model. In addition, this is the first work, classical or quantum, that considers the presence of Byzantine servers together with eavesdroppers in the same system model (static or dynamic eavesdroppers). Another layer of difficulty that we handle in this work stems from the \emph{symmetric} privacy requirement in the presence of Byzantine servers, which by itself has never been studied before in classical or quantum variations; here the Byzantine servers may attempt to leak information about undesired messages to the user, which is not allowed in SPIR.
\end{abstract}

\section{Introduction} 

\subsection{Problem Introduction and Our Contributions}
Recently, the quantum private information retrieval (QPIR) problem was introduced \cite{qpir}. In that setting, there are $N$ servers that share an entangled state. A user, who wishes to retrieve a message privately from $K$ messages, sends queries over classical channels to each server independently. Upon receiving the queries, the servers, who share a globally known entangled state, apply quantum operations on their individual quantum subsystems and send them to the user over separate quantum channels. It was shown in \cite{qpir} that the capacity for the quantum variation with symmetric privacy (QSPIR) is 1 when $N=2$. In \cite{qtpir}, the $T$-colluding QPIR is investigated and it was shown that the rate is $R= \min\{1, 2(\frac{N-T}{N})\}$. More recently, a new mathematical abstraction has been developed to model the system with $N$ entangled transmitters in a classical quantum setting that transmits over $N$ noise-free separate quantum channels to one receiver where the encoding of the classical information is done using unitary Pauli operations and the decoding is done using their corresponding projective value measurements (PVMs) \cite{nsumbox}. This abstraction is coined as \emph{$N$-sum box abstraction}. It was shown in \cite{nsumbox} that the quantum variation of the $X$-secure, $T$-colluding PIR (QXTPIR) achieves a rate equal to $R_Q = \min\{1, 2(\frac{N-X-T}{N})\}$.

This paper first investigates how to merge the dynamic eavesdropper threat with unresponsive servers in a single framework. Afterward, a new model that is not studied in classical or quantum symmetric PIR (SPIR) settings is investigated. In particular, we analyze the interactions between Byzantine servers, as defined in \cite{byzantine_tpir}, and eavesdroppers in a QSPIR system. We consider two eavesdropper models. In the first kind, the eavesdropper is \emph{static}, where the uplinks and downlinks listened to are identical \cite{sun_eaves, banawan_eaves}. In the second kind, the eavesdropper is \emph{dynamic}, where the eavesdropped links can be different for the uplink and downlink directions. The latter is considered only in \cite{our_journal}. We find the rate regions for both cases for the general case of the quantum $U$-unresponsive, $T$-colluding, and $X$-secure SPIR model. This is the first work that considers this kind of Byzantine behavior under symmetric privacy constraints in a classical or quantum setting. In addition, this is the first work that investigates this kind of Byzantine server with static or dynamic eavesdroppers.

\subsection{Related Work}
Private information retrieval (PIR) is the problem that considers the retrieval of a message by a user, out of $K$ messages stored in a replicated manner in $N$ servers, without revealing the required message index to any individual server \cite{chor}. The main performance metric for PIR is the ratio between the number of required message symbols and the total number of symbols downloaded, coined as the PIR rate. In \cite{c_pir}, it was shown that the optimal rate, i.e., the capacity, of PIR is $\left(1+\frac{1}{N}+\frac{1}{N^2}+\cdots+\frac{1}{N^{K-1}}\right)^{-1}$. In \cite{c_spir}, another requirement was added, coined symmetric privacy. In symmetric PIR (SPIR), the servers want the other, non-required, messages to be hidden from the user. It was shown that the capacity of SPIR is $1-\frac{1}{N}$ \cite{c_spir}. Another variation was formulated in \cite{c_tpir}, coined $T$-colluding PIR (TPIR), where any $T$ out of $N$ servers can collude to share their queries received from the user to break the user privacy. It was shown that the capacity of TPIR is $\left(1+\frac{T}{N}+\frac{T^2}{N^2}+\ldots+\frac{T^{K-1}}{N^{K-1}}\right)^{-1}$. In \cite{tspir_mdscoded}, the general problem of TSPIR was solved and the capacity is given by $1-\frac{T}{N}$. Another version for collusion was defined in \cite{arbitrarycollusion} for arbitrary collusion patterns. In \cite{byzantine_tpir}, the Byzantine version of TPIR problem (BTPIR) was formulated, where any $B$ out of $N$ servers can act as Byzantine servers and respond to the user queries with arbitrarily errored answers. It was shown that the capacity of BTPIR is $\left(\frac{N-2B}{N}\right)\left(1+\frac{T}{N-2B}+\ldots+\frac{T^{K-1}}{{N-2B}^{K-1}}\right)^{-1}$. In \cite{c_TESPIR}, the eavesdropped version of TSPIR was defined including active eavesdroppers. It was shown that if there are $E$ passive eavesdropped links and $A$ active eavesdropped links, then the capacity for ETSPIR is $1-\frac{2A+\max(E,T)}{N}$. Subsequently, the capacity of the ETPIR, i.e., with no symmetry requirement, was found in \cite{sun_eaves}. Another interesting model for eavesdroppers can be found in \cite{banawan_eaves}, where a wiretap channel viewpoint is used to model the eavesdroppers. 

It is important to note that there is a significant difference between the servers being Byzantine and the eavesdroppers being active. In the simplest sense, the Byzantine servers can help the eavesdropper get some information about the messages and the queries. However, an active eavesdropper can act on the answers being transmitted by the servers to cause errors at the user side. In addition, in the symmetric privacy setting, there is a shared randomness among all servers, which causes more issues when the servers themselves are Byzantine. Another variant was studied in \cite{unresponsive_byzantine_1}, where any $U$ servers can be unresponsive, and $B$ can be Byzantine (UBTPIR). The achievable rate is given by $1- \frac{N-U-T-2B}{N}$, and is conjectured to remain the same in the symmetric privacy setting. The main difference between that work and our work, aside from the presence of an eavesdropper, is that in \cite{unresponsive_byzantine_1} the Byzantine servers send random symbols independently of the other servers, the storage, and the queries. It is clear that our model here is more strict since we assume possible cooperation between the Byzantine servers. In the collaborative Byzantine servers setting, if any Byzantine server decides to send a message symbol randomly, the symmetric privacy requirement is violated. The latest model of PIR is $X$-secure TPIR (XTPIR) in \cite{first_xsecure}. The exact capacity of this setting is still an open problem. The highest achievable rate to date is $1-\frac{X+T}{N}$ \cite{csa}. Generally, for the case of $X$-secure, $B$-Byzantine, $U$-unresponsive, $T$-colluding PIR (XBUTPIR), the state of the art rate is given by \cite{jafar_byzantine}. More variations and results can be found in \cite{banawan_multimessage_pir, banawan_pir_mdscoded, batuhan_hetero, ChaoTian, codedstorage_adversary_tpir, grpahbased_pir, wei_banawan_cache_pir, nomeirasymmetric, uncoded_constrainedstorage_pir, ulukusPIRLC, semantic_pir, salim_singleserver_pir, tpir_sideinfo, Salim_CodedPIR}, in addition to the newly developed approaches using algebraic geometric codes in PIR in \cite{AG_1, AG_2}.

As for the QPIR literature, in \cite{our_journal,our_quantum_first} the eavesdropper variation was studied and the rate regions were found. In \cite{jafar_quantum_unresponsive}, the quantum unresponsive servers variation was analyzed and the rate regions were given. Finally, the quantum Byzantine servers version was analyzed and the rate regions were found in \cite{byzantine_1}. More applications for the new $N$-sum box abstraction can be found in \cite{lu_product, yao_capacity_MAC, yao_inverted}. 

\section{Problem Formulation}
We consider a system consisting of $N$ servers and $K$ messages. The messages are stored classically within the servers. Each message $W_i$ consists of $L$ symbols chosen uniformly and independently at random from the field $\mathbb{F}_q$, i.e.,
\begin{align}
    H(W_i) = & L, \\
    H(W_{[K]}) = & KL,
\end{align}
where $H(\cdot)$ is the Shannon entropy with base $q$ \cite{coverthomas}.

Let $S_n$ denote the storage at the $n$th server. Any $\mathcal{X} \subset [N] $ servers can communicate to attempt to decode the stored content where $|\mathcal{X}| \leq X$. Thus, the message content must be hidden from any $X$ communicating servers, i.e.,
\begin{align}
    I(W_{[K]};S_{\mathcal{X}}) = 0.
\end{align}

A user chooses $\theta$ uniformly at random from $[K]$ to retrieve the $\theta$th message. The user sends the query $Q_n^{[\theta]}$ over a classical channel to the $n$th server. Any $\mathcal{T} \subset [N] $ servers can collude to attempt to decode the required message index $\theta$, where $ |\mathcal{T}| \leq T$. Thus, the queries must be secure against any $\mathcal{T}$ colluding servers, i.e.,
\begin{align}
    I(Q_{\mathcal{T}}; \theta) = 0.
\end{align}

The servers share a quantum system $\mathcal{A}^{[0]} = \mathcal{A}_1^{[0]}\ldots\mathcal{A}_N^{[0]}$  represented by the quantum state $\rho$ prior to the initiation of the scheme. The honest but curious servers encode their answers over their own part of the state $\rho_n = {tr}_{i \in [N]\setminus n}\rho$ and reply back over their separate quantum channels to the user using their individual stored symbols $S_n$, their received queries $Q_n^{[\theta]}$, and the shared common randomness among all the servers $Z'$, i.e., the honest servers generate the classical answers such that
\begin{align}
    &H(A_n|S_n,Q_n, Z') =  0,
\end{align}
and encode them over their separate quantum systems, i.e.,
\begin{align}
\mathcal{A}_n =  \text{Enc}_n(A_n, \mathcal{A}_n^{[0]}),
\end{align}
where $\text{Enc}_n(\cdot,\cdot)$ is the quantum encoder at the $n$th server.

In addition, the Byzantine servers generate classical answers as an arbitrary function of their received queries, storage, shared common randomness, and an arbitrary random variable $\Delta_n$, i.e., 
\begin{align}
    H(A_n|S_n,Q_n,Z',\Delta_n) = 0, 
\end{align}
and 
\begin{align}
    \mathcal{A}_n = \text{Enc}'_n(A_n,\mathcal{A}_n^{[0]}),
\end{align}
where $\text{Enc}'_n(\cdot,.\cdot)$ is any arbitrary quantum encoder.

\begin{remark}
    As mentioned earlier, here we consider Byzantine servers that send answers not just based on an independent random variable but can be any arbitrary function of the storage, queries, masking random variables, and an arbitrary random variable.
\end{remark}

Upon receiving all the answers, the user must be able to decode the required message $W_\theta$ { after applying the quantum measurement operations on the received quantum answers}, i.e., 
{
\begin{align}
    H(W_{\theta}| Y, Q_{[N]}) = 0,
\end{align}
where $Y$ is the result of the measurements on the received quantum answers.
}

In addition, to achieve symmetric privacy, the user cannot get any information about messages other than the desired message $W_\theta$, i.e.,
\begin{align}\label{symmetric_privacy}
    S(W_{[K] \setminus \theta}; \mathcal{A}_{[N]}, Q_{[N]})=0.
\end{align}
Note that \eqref{symmetric_privacy} includes both honest and Byzantine servers. 

Furthermore, there are $E$ eavesdropped uplinks and $E$ eavesdropped downlinks where the passive eavesdropper can listen to any set of queries and answers. Both the index and the messages should be secure against the eavesdropper, i.e.,
\begin{align}\label{eaves_eq}
    S(W_{[K]}, \theta; Q_{\mathcal{E}_1}, \mathcal{A}_{\mathcal{E}_2})=0,
\end{align}
where $\mathcal{E}_1, \mathcal{E}_2 \subset [N]$, and  $|\mathcal{E}_1|, |\mathcal{E}_2|\leq E$.

Prior to the presentation of our main results in the next section, we make the following important remarks. 

\begin{remark}
    As stated earlier, in the PIR and QPIR literature, two different kinds of eavesdropper models have been considered. In the first kind, the eavesdropper is \emph{static}, which means that the uplinks and downlinks listened to are identical, i.e., $\mathcal{E}_1 = \mathcal{E}_2$ in \eqref{eaves_eq}. In the second kind, the eavesdropper is \emph{dynamic}, which means that the eavesdropped links can be different for the uplink and the downlink. The latter is considered only in \cite{our_journal} so far in the literature. However, as analyzed in \cite{our_journal}, the static or dynamic eavesdroppers are essentially the same when acting alone with no other harmful entity.
\end{remark}

\begin{remark}
    If there are no eavesdroppers in the system, the Byzantine servers have fewer options for actions to harm the system, since the scheme needs only to protect for the user and the databases, i.e., symmetric, privacy. However, when there is a dynamic eavesdropper in the system, the Byzantine servers can also relay part of the queries they have received. As a simple example, consider the case that the eavesdropper listens to the servers $\mathcal{E}_1$ in the uplink and the servers $\mathcal{E}_2$ in the downlink, where $\mathcal{B} \subset \mathcal{E}_2$, and $\mathcal{E}_1 \cap \mathcal{E}_2 = \emptyset$, with $|\mathcal{E}_1|, |\mathcal{E}_2| \leq E$. In this case, if the queries are protected against $E$ links only and if the Byzantine servers decide to re-transmit part of the queries, then the scheme will be compromised.
\end{remark}

\section{Main Results}

\begin{theorem}\label{eaves_only_thm}
    For a QSPIR system with $N$ databases, where $X$ of them can communicate to decode messages, $|\mathcal{E}_1|\leq E$ of the uplinks and $|\mathcal{E}_2| \leq E$  of the downlinks are tapped by a passive eavesdropper, $U$ of the databases are unresponsive, and $T$ of them are colluding to identify user-required message index, the following rate is achievable with symmetric privacy, 
    \begin{align}
        R_Q = \begin{cases}
            2\big(\frac{N-X-M-U}{N}\big), & N-U> X+M \geq \frac{N}{2}, \ E \leq 2X+2M-N, \\
            2\big(\frac{N-X-M-U-\frac{\delta}{2}}{N}\big), & N-U -\frac{\delta}{2}> X+M \geq \frac{N}{2}, \ E > 2X+2M-N, \\
            \frac{N-2U-E}{N}, & X+M < \frac{N}{2}, \ U+E < X+M, \ N> 2U+E,\\
            \frac{N-X-U-M}{N}, & X+M < \frac{N}{2}, \ U+E > X+M, \ N >X+M+U,
        \end{cases}
    \end{align}
    where $M = \max\{E,T\}$, and $\delta = N+E-2X-2M$.
\end{theorem}

\begin{theorem}\label{byzantine_static_eaves_thm}
    For a QSPIR system with $N$ databases, where $X$ of them can communicate to decode messages, $B$ are Byzantine, $E$ links are tapped by passive eavesdroppers that can listen to both uplink and downlink communications (static eavesdropper), $U$ of the databases are unresponsive, and $T$ of them are colluding to identify user-required message index, the following rate is achievable with symmetric privacy, 
    \begin{align}
        R_Q = \begin{cases}
            2\big(\frac{N-H-M-3B-U}{N}\big), & N-U > H+M+B \geq \frac{N}{2}, \ E \leq 2H+2M+2B-N, \\
            2\big(\frac{N-H-M-3B-U-\frac{\delta}{2}}{N}\big), &N-U-\frac{\delta}{2}> H+M+B \geq \frac{N}{2}, \ E > 2H+2M+2B-N, \\
            \frac{N-6B-2U-E}{N}, & H+M+B < \frac{N}{2}, \ 2B+U+E < H+M+B, \ N > 6B+2U+E,\\
            \frac{N-H-3B-U-M}{N}, & H+M+B < \frac{N}{2}, \ 2B+U+E > H+M+B, \ N> H+3B+U+M,
        \end{cases}
    \end{align}
where $M = \max\{E,T\}$, $H = \max\{X,B\}$,  and $\delta = N+E -2H-2M-2B$.
\end{theorem}

\begin{theorem}\label{byzantine_eaves_thm}
    For a QSPIR system with $N$ databases, where $X$ of them can communicate to decode messages, $B$ are Byzantine, $|\mathcal{E}_1|\leq E$ of the uplinks and $|\mathcal{E}_2| \leq E$ of the downlinks are tapped by a passive eavesdropper, $U$ of the databases are unresponsive, and $T$ of them are colluding to identify user-required message index, the following rate is achievable with symmetric privacy,
    \begin{align}
        R_Q = \begin{cases}
            2\big(\frac{N-H-M-3B-U}{N}\big) , & N-U> H+M+B \geq \frac{N}{2}, \ E \leq 2H+2M+2B-N, \\
            2\big(\frac{N-H-M-3B-U-\frac{\delta}{2}}{N}\big), & N-U-\frac{\delta}{2} > H+M+B \geq \frac{N}{2}, \ E > 2H+2M+2B-N, \\
            \frac{N-6B-2U-E}{N}, & H+M+B < \frac{N}{2}, \ 2B+U+E < H+M+B, \ N > 6B+2U+E,\\
            \frac{N-H-3B-U-M}{N}, & H+M+B < \frac{N}{2}, \ 2B+U+E > H+M+B, \ N > H+3B +U+2M,
        \end{cases}
    \end{align}
where $M = \max\{E+B,T\}$, $H = \max\{X,B\}$, and $\delta = N+E-2H-2M-2B$.
\end{theorem}

\begin{remark}
    It is clear that there is a significant additional threat imposed on the system when we have both dynamic eavesdroppers and Byzantine servers in the system, while providing symmetric privacy. Further, allowing the eavesdropper to jump from any $E$ links in the upload phase to another $E$ links in the retrieval phase makes the setting more complex. This can be observed by comparing the rates in Theorem~\ref{eaves_only_thm} and Theorem~\ref{byzantine_eaves_thm}.
\end{remark}

\begin{remark}
    Note the difference in the effect of a dynamic eavesdropper in Theorem~\ref{byzantine_eaves_thm} compared to a static eavesdropper in Theorem~\ref{byzantine_static_eaves_thm} in terms of the achievable rate. The main difference lies in the fact that relaying the queries again by the Byzantine servers is not useful in the case of the static eavesdropper since they were able to fetch them during the uplink communication from the user to the servers.
\end{remark}

\begin{remark}
    In the achievable schemes sections, we present the achievable schemes for Theorem~\ref{eaves_only_thm} and Theorem~\ref{byzantine_eaves_thm}. The only modification required for the achievable scheme in Theorem~\ref{byzantine_static_eaves_thm} compared to that in Theorem~\ref{byzantine_eaves_thm} is to hide queries within a smaller vector subspace.
\end{remark}

\begin{remark}
    In Section~\ref{proofs}, we prove the properties of the main threat model given in Theorem~\ref{byzantine_eaves_thm}, since other variations can be derived from it as they have fewer restrictions, i.e., $B=0$ for proving properties of Theorem~\ref{eaves_only_thm}, and $\mathcal{E}_1 = \mathcal{E}_2$ for Theorem~\ref{byzantine_static_eaves_thm}.
\end{remark}

\begin{remark}
    In Theorem~\ref{byzantine_static_eaves_thm}, the security parameter changes from $X$ to $H = \max\{X,B\}$. Intuitively, the main reason behind this is to make sure that, if all $B$ Byzantine servers send the $m$th symbol for the $\kappa$th message $W_{\kappa,m}$ to the user, the user will not be able to decode it.
\end{remark}

\begin{remark}
    In Theorem~\ref{byzantine_eaves_thm}, the colluding parameter $M = \max\{E,T\}$ is modified to $M = \max\{E+B,T\}$. Logically, this is done to avoid the issue of Byzantine servers relaying the queries back and the potential of eavesdropping on those $B$ links by the dynamic eavesdropper.
\end{remark}

\begin{remark}
    In both Theorem~\ref{byzantine_static_eaves_thm} and Theorem~\ref{byzantine_eaves_thm}, a $3B$ factor appears instead of the usual $2B$ factor due to $B$ Byzantine servers. The main reason for this is to hide the masking used for symmetric privacy from the collection of Byzantine servers. That is, if the masking variables are known, the Byzantine servers can help the user to decode the interference terms and leak information about the other messages.
\end{remark}

\section{Preliminaries}

The following notation is adopted in this paper to ease reading: For matrices, math typewriter font is used, e.g., $\mathtt{A}$. For column vectors, bold math font is used, e.g., $\bm{a}$. 

\subsection{Quantum Information Theory}
We provide some important definitions for completeness. These can be found in introductory quantum information processing books, e.g., \cite{nielsen-chuang, hayashibook}.

\begin{definition}[Quantum density matrices]
    For a quantum system $A$, that can be in the state $\ket{\psi_j}$ with probability $p_j$, the quantum density matrix $\rho_A$ is defined as,
    \begin{align}
        \mathtt{\rho}_A = \sum_j p_j \ket{\psi_j}\bra{\psi_j},
    \end{align}
with $p_j \geq 0$ and $\sum_j p_j =1$.
\end{definition}

\begin{definition}[Von Neumann entropy]
    For the density matrix $\rho$, the Von Neumann entropy is defined as,
    \begin{align}
        S(\mathtt{\rho})= -tr(\rho \log\rho)=H(\bm{\Lambda}),
    \end{align}
    where $tr(\cdot)$ is the trace operator, $\bm{\Lambda}$ are the eigenvalues of $\rho$, and $H(\cdot)$ is the Shannon entropy. For a quantum system $A$ with density matrix $\rho_A$, we define $S(A) = S(\rho_A)$.
\end{definition}

\begin{definition}[Quantum conditional entropy]
    The conditional entropy of a quantum system $A$ with respect to a system $B$ is defined as,
    \begin{align}
        S(A|B) = S(A,B)-S(B).
    \end{align}
\end{definition}

\begin{definition}[Quantum mutual information]
    The quantum mutual information between two quantum systems $A$ and $B$ is defined as,
    \begin{align}
        S(A;B)&=S(A)+S(B)-S(A,B)\\
        &=S(A)-S(A|B).
    \end{align}
\end{definition}

\begin{definition}[Quantum operation]
    A quantum operation $\mathcal{E}$ is a linear, completely-positive, and trace-preserving map from the set of all density matrices to itself.
\end{definition} 

\begin{definition}[Kraus representation] \label{kraus}
    Any quantum operation $\mathcal{E}$ acting on a quantum state $\mathtt{\rho}$ can be written in the form,
    \begin{align}
        \mathcal{E}(\rho) = \sum_i \mathtt{M}_i\mathtt{\rho} \mathtt{M}_i^{\dagger},
    \end{align}
    where $\sum_i \mathtt{M}_i^\dagger \mathtt{M}_i= \mathtt{I}$.
\end{definition}

\subsection{$N$-Sum Box Abstraction}
The encoding and decoding structure using the $N$-sum box abstraction introduced recently in \cite{nsumbox} is defined as follows. In the encoding stage, the databases use generalized Pauli operators $\mathsf{X}(a) = \sum_{j=0}^{q-1} \ket{j+a}\bra{j}$, and  $\mathsf{Z}(a) = \sum_{j=0}^{q-1} \omega^{tr(aj)} \ket{j}\bra{j}$, where $q=p^r$ with $p$ being a prime number, $a \in \mathbb{F}_q$ and $\omega = \exp(2\pi i /p)$. 

In the decoding stage, the user applies projection-valued measurement (PVM) defined on the quotient space of the stabilizer group $\mathcal{L}(\mathcal{V})$ defined by,
\begin{align}
    \mathcal{L}(\mathcal{V}) = \{c_{\bm{v}} \mathtt{W}(\bm{v}) : \bm{v} \in \mathcal{V} \},
\end{align}
where $\mathcal{V}$ is a self-orthogonal subspace in $\mathbb{F}_q ^{2N}$, 
\begin{align}
     \mathtt{W}(\bm{v}) = \mathsf{X}(v_1) \mathsf{Z}(v_{N+1}) \otimes \ldots \otimes \mathsf{X}(v_N) \mathsf{Z}(v_{2N}),
\end{align}
and $c_{\bm{v}} \in \mathbb{C}$ is chosen such that $\mathcal{L}(\mathcal{V})$ is an Abelian subgroup of $HW_{q}^N$  with $c_{\bm{v}}\mathtt{I}_{q^N}$ being an element of the stabilizer group only with $c_{v}=1$, where $HW_{q}^N$ is the Heisenberg-Weyl group defined as follows,
\begin{align}
    HW_{q}^N = \{ c \mathtt{W}(\bm{s}) : \bm{s} \in \mathbb{F}_q^{2N}, c \in \mathbb{C} \setminus \{0\} \}.
\end{align}

\begin{theorem}[Thm.~1  in \cite{nsumbox}]\label{feasible_nsum_1}
    Let $\mathtt{G}$ be a $2N\times N$ SSO matrix, i.e., $\mathtt{G^tJG}=0$, where $\mathtt{J} = \begin{bmatrix}
        0 &\mathtt{I}_N\\
        -\mathtt{I}_N&0
    \end{bmatrix}$, and $\mathtt{H}$ be a $2N\times N$ matrix with full column rank. If $[\mathtt{G} ~ \mathtt{H}]$ is a full-rank matrix, then $\mathtt{G}' = [0_N ~ \mathtt{I}_N ] [\mathtt{G} ~ \mathtt{H} ]^{-1}$ is a feasible $N$-sum box transfer matrix. 
\end{theorem}

\begin{definition}[Dual QCSA matrices \cite{qcsa}]\label{dual_def}
    The matrices $\mathtt{H}^{\bm{u}}_N$ and $\mathtt{H}^{\bm{v}}_N$ are defined as $\mathtt{H}^{\bm{u}}_N =\mathtt{QCSA}_{N\times N}(\bm{f}_{[L]},\bm{\alpha}_{[N]},\bm{u}_{[N]})$ and $\mathtt{H}^{\bm{v}}_N =\mathtt{QCSA}_{N\times N}(\bm{f}_{[L]},\bm{\alpha}_{[N]},\bm{v}_{[N]})$. Then, $\mathtt{H}^{\bm{u}}_N$, and $\mathtt{H}^{\bm{v}}_N$ are dual QCSA matrices if,
    \begin{enumerate}
        \item $u_1,\ldots,u_N$ are non-zero,
        \item $u_1,\ldots,u_N$ are distinct,
        \item for each $v_j,~j\in[N]$,
        \begin{align}
            v_j = \frac{1}{u_j} \Bigg(\prod_{i=1 \atop i\neq j}^N(\alpha_j-\alpha_i)\Bigg)^{-1},
        \end{align}
    \end{enumerate}
    where 
    \begin{align}
    \mathtt{QCSA}_{N,L, N}(\bm{f}_{[L]},\bm{\alpha}_{[N]},\bm{\beta}_{[N]})=\diag(\bm{\beta}_{[N]}) \mathtt{CSA}_{N,L, N}(\bm{f}_{[L]},\bm{\alpha}_{[N]}).
   \end{align}
\end{definition}

\begin{theorem}[Thm.~1 in \cite{qcsa}]\label{feasile_nsum_2}
    For any dual QCSA matrices $\mathtt{H}^{\bm{u}}_N$ and $\mathtt{H}^{\bm{v}}_N$, there exists a feasible $N$-sum box transfer matrix $\mathtt{G}(\bm{u},\bm{v})$ of size $N \times 2N$ given by,
    \begin{align}\label{tx-rx releation}
        \mathtt{G}(\bm{u},\bm{v}) = &\begin{bmatrix}
            \mathtt{I}_L&0_{L\times\nu} & 0 & 0 & 0 & 0\\
            0 & 0 & \mathtt{I}_{\mu-L} & 0 & 0 & 0\\ 0 & 0 & 0 & \mathtt{I}_L & 0_{L\times\mu} & 0\\
            0 & 0 & 0 & 0 & 0 & \mathtt{I}_{\nu-L}
            \end{bmatrix} \begin{bmatrix}
            \mathtt{H}^{\bm{u}}_N & 0\\
            0& \mathtt{H}^{\bm{v}}_N
        \end{bmatrix}^{-1},
    \end{align}
    where $\nu = \lceil N/2\rceil$ and $\mu = \lfloor N/2\rfloor$.
\end{theorem}

\section{Achievable Scheme for QXEUTSPIR}
In this section and the following section, we follow the same encoding and decoding procedure for the $N$-sum box. In the naming of the subsections, we include the main restrictions for each regime, and trim the ones that basically state the requirement of having sufficiently many servers for the scheme to work.

\subsection{Regime 1: $X+M \geq \frac{N}{2}$ and $ E \leq 2X+2M -N$}
Let $L = N-X-M-U$ and the storage at the $n$th database, $\bm{S}_n(i)$, $i \in [2]$, is
\begin{align}\label{classical_storage}
    \bm{S}_n(i)=\begin{bmatrix}
        \bm{W}_{\cdot,1}(i) + (f_1-\alpha_n)\bm{R}_{11}(i)+(f_1-\alpha_n)^2\bm{R}_{12}(i)+\cdots+(f_1-\alpha_n)^X\bm{R}_{1X}(i)\\
        \bm{W}_{\cdot,2}(i) + (f_2-\alpha_n)\bm{R}_{21}(i)+(f_2-\alpha_n)^2\bm{R}_{22}(i)+\cdots+(f_2-\alpha_n)^X\bm{R}_{2X}(i)\\
        \vdots\\
        \bm{W}_{\cdot,L}(i) + (f_L-\alpha_n)\bm{R}_{L1}(i)+(f_L-\alpha_n)^2\bm{R}_{L2}(i)+\cdots+(f_L-\alpha_n)^X\bm{R}_{LX}(i)\\
    \end{bmatrix},
\end{align}
where $\bm{W}_{\cdot,j}(i)=[W_{1,j}(i),\ldots,W_{K,j}(i)]^t$ is a vector representing the $j$th dit of all $K$ messages, with $W_{k,j}$ being the $j$th bit of message $k$ for the $i$th instance, $\bm{R}_{kj}(i)$ are uniform independent random vectors with the same dimensions as $\bm{W}_{\cdot,j}(i), ~ j \in [1:L]$, $\alpha_{\ell}, f_j \in \mathbb{F}_q$ are distinct where $\ell=1,\ldots,N$ and $j=1,\ldots,L$, and $t$ denotes the transpose operator.

To retrieve the $\theta$th message, the user sends the following queries to the $n$th server
\begin{align}\label{classical_queries}
    \bm{Q}_n^{[\theta]}=\begin{bmatrix}
        \frac{1}{f_1-\alpha_n}\left(\bm{e}_{\theta}+(f_1-\alpha_n)\bm{Z}_{11}+\cdots +(f_1-\alpha_n)^M \bm{Z}_{1M}\right)\\
        \vdots\\
        \frac{1}{f_L-\alpha_n}\left(\bm{e}_{\theta}+(f_L-\alpha_n)\bm{Z}_{L1}+\cdots+ (f_L-\alpha_n)^M \bm{Z}_{LM}\right)
    \end{bmatrix},
\end{align}
where $\bm{e}_{\theta}$ is a vector of length $K$ with $1$ in the $\theta$th index and zero otherwise, and $\bm{Z}_{ij}$ are uniform independent random vectors of length $K$ each, chosen by the user.

Let $\mathcal{U}= \{n_1,\ldots,n_U\}$ denote the set of unresponsive servers. The answer from the $n$th server, where $n \in [N]\setminus \mathcal{U}$ is given by
\begin{align}\label{s_1_r}
    A_n^{[\theta]}(i) = \bm{S}_n(i)^t \bm{Q}_n^{[\theta]}+ \sum_{j=0}^{X+M-1}\alpha_n^j Z_j'(i), 
\end{align}
where $Z_j'(i)$ are generated uniformly at random and shared among servers. Note that the answers from the $N-U$ servers, $i_1, \ldots, i_{N-U}$, for one instance only of the classical scheme, can be written compactly as 
\begin{align}
    \begin{bmatrix}
        A_{i_1}^{[\theta]}\\
        \vdots\\
        A_{i_{N-U}}^{[\theta]}
    \end{bmatrix} &=\underbrace{\begin{bmatrix}
        \frac{1}{f_1-\alpha_{i_1}} &\cdots&\frac{1}{f_L-\alpha_{i_1}}&1&\alpha_{i_1}&\cdots&\alpha_{i_1}^{N-U-L-1}\\
        \frac{1}{f_1-\alpha_{i_2}}&\cdots&\frac{1}{f_L-\alpha_{i_2}}&1&\alpha_{i_2}&\cdots&\alpha_{i_2}^{N-U-L-1}\\
        \vdots&\vdots&\vdots&\vdots&\vdots&\vdots&\vdots\\
        \frac{1}{f_1-\alpha_{i_{N-U}}}&\cdots&\frac{1}{f_L-\alpha_{i_{N-U}}}&1&\alpha_{N}&\cdots&\alpha_{i_{N-U}}^{N-U-L-1}
    \end{bmatrix}}_{\mathtt{CSA}_{N-U, L,N-U}}\begin{bmatrix}
        \bm{W}_{\theta}\\
        \bm{I}_{[X+M]}
    \end{bmatrix}  \\
    &=\mathtt{CSA}_{N-U, L,N-U} \bm{x},
\end{align}
where 
\begin{align}
    \bm{x} = \begin{bmatrix}
        \bm{W}_{\theta}\\
        \bm{I}_{[X+M]}
    \end{bmatrix},
\end{align}
where $\bm{I}_{[X+M]}$ is the interference vector which is discarded after decoding. In the classical case, the user does not receive any symbols from the unresponsive servers which can be identified easily in the classical regime. However, in the quantum setting the user has to apply initial measurements to know the unresponsive servers. Thus, in the quantum setting, unresponsiveness can be thought of as an erasure error instead of non-received qudits as shown in \cite{jafar_quantum_unresponsive}. The main reason for this is that the user can apply initial measurements to know which servers responded without affecting the remaining quantum state \cite{gottesman_book}. Based on the initial measurements, and the unresponsive servers, the actual measurements for decoding the message symbols can then be applied. As presented in \cite{qcsa}, in the quantum setting two instances of the classical scheme can be transmitted in one instance using the quantum scheme when the servers encode the symbols on the entangled shared state whenever possible. Thus, the answers of the responsive servers can be written as 
\begin{align}
    \mathtt{X}(u_nA_n(1))\mathtt{Z}(v_nA_n(2))\rho_n\left(\mathtt{X}(u_nA_n(1))\mathtt{Z}(v_nA_n(2))\right)^\dagger, \quad n \in [N] \setminus \mathcal{U},
\end{align}
{where $\rho_n$ is the density matrix for the quantum system $\mathcal{A}_n$}.

As for the unresponsive servers in the quantum setting, the answers for $ n \in \mathcal{U}$ can be written as
\begin{align}
   &\mathtt{X}(u_n\Delta'_n(1))\mathtt{Z}(v_n\Delta'_n(2)) \mathtt{X}(u_nA_n(1))\mathtt{Z}(v_nA_n(2)){ \rho_n}\left(\mathtt{X}(u_n\Delta'_n(1))\mathtt{Z}(v_n\Delta'_n(2)) \mathtt{X}(u_nA_n(1))\mathtt{Z}(v_nA_n(2))\right)^\dagger \nonumber\\
   & = \mathtt{X}(u_n(A_n(1)+\Delta'_n(1)))\mathtt{Z}(v_n(A_n(2)+\Delta'_n(2))) { \rho_n}\left(\mathtt{X}(u_n(A_n(1)+\Delta'_n(1)))\mathtt{Z}(v_n(A_n(2)+\Delta'_n(2)))\right)^\dagger,
\end{align}
where $\mathcal{A}_n$ is the subsystem available at the $n$th database, $\Delta'_n$ is generated uniformly at random, and $A_n(i)$ is as in \eqref{s_1_r}.
 
Now, for one instance of the quantum scheme, the collective answers  can be written as follows
\begin{align}
\bm{A}^{[\theta]}=& \begin{bmatrix}
     \mathtt{\Lambda}_1(:,1:\lceil\frac{N}{2}\rceil)& 0&\mathtt{\Gamma}_1&0&\mathtt{\Lambda}_1(:,a_1 :X+M) &0&\mathtt{\Omega}&0\\
    0&\mathtt{\Lambda}_2(:,1:\lfloor\frac{N}{2}\rfloor)&0&\mathtt{\Gamma}_2&0&\mathtt{\Lambda}_2(:,a_2 :X+M)&0&\mathtt{\Omega}
    \end{bmatrix}\times \nonumber\\
    &\quad\begin{bmatrix}
        \bm{I}_1(1)\\
        \bm{I}_1(2)\\
        \bm{W}_{\theta}(1)\\
        \bm{W}_{\theta}(2)\\
        \bm{I}_2(1)\\ \bm{I}_2(2)\\
        \bm{\Delta}_U'(1)\\ \bm{\Delta}_U'(2)
    \end{bmatrix},
\end{align}
where $\mathtt{\Lambda}_1 = \mathtt{GRS}_{N,X+M}^{\bm{\alpha}_{[N]},\bm{u}_{[N]}}$, $\mathtt{\Lambda}_2 = \mathtt{GRS}_{N,X+M}^{\bm{\alpha}_{[N]},v_{[N]}}$, $\mathtt{\Gamma}_1 = \mathtt{GC}_{N,L}^{\bm{\alpha}_{[N]},\bm{u}_{[N]},\bm{f}_{[L]}}$, $\mathtt{\Gamma}_2 = \mathtt{GC}_{N,L}^{\bm{\alpha}_{[N]},v_{[N]},\bm{f}_{[L]}}$, $a_1 = \lceil\frac{N}{2}\rceil+1$, $a_2 = \lfloor\frac{N}{2}\rfloor+1$, and $\mathtt{\Omega} = [\bm{e}_{n_1}, \bm{e}_{n_2}, \ldots, \bm{e}_{n_U}]$, where $\bm{e}_{n_i}$ is the unit vector of length $N$ and $0$ in all locations except the $n_i$th index, and $\mathcal{U} = \{n_1, \ldots, n_U\}$.

In addition,
\begin{align}
    \mathtt{GRS}_{N,X+M}^{\bm{\alpha}_{[N]},\bm{u}_{[N]}} = \begin{bmatrix}
        u_1& u_1 \alpha_1& \ldots & u_1\alpha_1^{X+M-1}\\
        u_2& u_2 \alpha_2& \ldots& u_2\alpha_2^{X+M-1}\\
        \vdots&\vdots&\ddots&\vdots\\
        u_N& u_N \alpha_N& \ldots& u_N\alpha_N^{X+M-1}
    \end{bmatrix},
\end{align}
and 
\begin{align}
    \mathtt{GC}_{N,L}^{\bm{\alpha}_{[N]},\bm{u}_{[N]},\bm{f}_{[L]}}=\begin{bmatrix}
        \frac{u_1}{f_1-\alpha_1}&\frac{u_1}{f_2-\alpha_1}& \ldots& \frac{u_1}{f_L-\alpha_1}\\
        \frac{u_2}{f_1-\alpha_2}&\frac{u_2}{f_2-\alpha_2}& \ldots& \frac{u_2}{f_L-\alpha_2}\\
        \vdots& \vdots& \ddots&\vdots\\
        \frac{u_N}{f_1-\alpha_N}&\frac{u_N}{f_2-\alpha_N}& \ldots& \frac{u_N}{f_L-\alpha_N}
    \end{bmatrix}.
\end{align}

Since there are $E \leq 2X+2M-N$ eavesdropped links, the servers can use a channel transmission matrix as follows
\begin{align}
    \mathtt{G}'(\bm{u},\bm{v}) = \begin{bmatrix}
        0_N& \mathtt{I}_N
    \end{bmatrix}\underbrace{\begin{bmatrix}
        \mathtt{I}_N&0&0\\
        0&\mathtt{V}_{2L+2X+2M-N}(b_{[2L+2X+2M-N]})& 0 \\
        0&0& \mathtt{I}_{2U}
    \end{bmatrix}
        \mathtt{M}^{-1}}_{\mathtt{M}_1^{-1}},
\end{align}
where
\begin{align}
    \mathtt{V}_{N}(\bm{b}_{[N]}) = \begin{bmatrix}
        1&b_1&\ldots&b_1^{N-1}\\
        \vdots & & & \vdots\\
        1&b_{N}&\ldots&b_{N}^{N-1}
    \end{bmatrix},
\end{align}
and
\begin{align}\label{main_transfer}
     \mathtt{M} = \begin{bmatrix}
     \mathtt{\Lambda}_1(:,1:\lceil\frac{N}{2}\rceil)& 0&\mathtt{\Gamma}_1&0&\mathtt{\Lambda}_1(:,a_1 :X+M) &0&\mathtt{\Omega}&0\\
    0&\mathtt{\Lambda}_2(:,1:\lfloor\frac{N}{2}\rfloor)&0&\mathtt{\Gamma}_2&0&\mathtt{\Lambda}_2(:,a_2 :X+M)&0&\mathtt{\Omega}
    \end{bmatrix}.
\end{align}

\begin{remark}
    The proof for the feasibility of the chosen transmission matrix $\mathtt{G}'(\bm{u},\bm{v})$ for this regime and the remaining regimes is given in Section~\ref{proofs}.
\end{remark}
 
Now, the received symbols by the user after the final measurements can be written as
\begin{align}
    \bm{y} &= \begin{bmatrix}
        0_N& \mathtt{I}_N
    \end{bmatrix}\begin{bmatrix}
        \mathtt{I}_N&0&0\\
        0&\mathtt{V}_{2L+2X+2M-N}(\bm{b}_{[2L+2X+2M-N]})& 0 \\
        0&0& \mathtt{I}_{2U}
    \end{bmatrix}\begin{bmatrix}
        \bm{I}_1(1)\\\bm{I}_1(2)\\\bm{W}_{\theta}(1)\\ \bm{W}_{\theta}(2)\\ \bm{I}_2(1)\\ \bm{I}_2(2)\\ \bm{\Delta}_U'(1)\\ \bm{\Delta}_U'(2)
    \end{bmatrix}\\
    &= \begin{bmatrix}
        \mathtt{V}_{2L+2X+2M-N}(b_{[2L+2X+2M-N]})\begin{bmatrix}
            \bm{W}_{\theta}(1)\\ \bm{W}_{\theta}(2)\\ \bm{I}_2(1)\\ \bm{I}_2(2)
        \end{bmatrix}\\
        \bm{\Delta}_U'(1)\\
        \bm{\Delta}_U'(2)
    \end{bmatrix} \label{s_1_e}.
\end{align}
Since the first submatrix in \eqref{s_1_e} is invertible, the user can decode the message symbols.

\subsection{Regime 2: $X+M \geq \frac{N}{2}$ and $E > 2X+2M-N $}
In this case, we develop a scheme that can use the interference symbols along with the noise variables introduced in the storage to prevent the eavesdropper from decoding the messages. As the number of eavesdropped links is greater than the number of interference symbols received by the user, we need to introduce new noise variables to compensate for the difference. Let $\delta = E -(2X+2M-N)$, $L_1=N-X-M-U-\delta$, and $L_2 =L_1+\delta$. Define $\bm{r}_{.,1},\ldots,\bm{r}_{.,\delta}$ to be uniform random vectors in $\mathbb{F}_q$. Then, the storage is given by 
\begin{align}
    \bm{S}_n(1) = \begin{bmatrix}
        \bm{r}_{\cdot,1}+ (f_1-\alpha_n)\bm{R}_{11}(1)+(f_{1}-\alpha_n)^2\bm{R}_{12}(1)+\ldots+(f_{1}-\alpha_n)^X\bm{R}_{1X}(1)\\
        \vdots\\
        \bm{r}_{\cdot,\delta}+ (f_{\delta}-\alpha_n)\bm{R}_{{\delta}1}(1)+(f_{\delta}-\alpha_n)^2\bm{R}_{{\delta}2}(1)+\ldots+(f_{\delta}-\alpha_n)^X\bm{R}_{{\delta}X}(1)\\
        \bm{W}_{\cdot,1} + (f_{\delta+1}-\alpha_n)\bm{R}_{\delta+1,1}(1)+(f_{\delta+1}-\alpha_n)^2\bm{R}_{\delta+1,2}(1)+\ldots+(f_{\delta+1}-\alpha_n)^X\bm{R}_{\delta+1,X}(1)\\
        \bm{W}_{\cdot,2} +(f_{\delta+2}-\alpha_n)\bm{R}_{\delta+2,1}(1)+(f_{\delta+2}-\alpha_n)^2\bm{R}_{\delta+2,2}(1)+\ldots+(f_{\delta+2}-\alpha_n)^X\bm{R}_{\delta+2,X}(1)\\
        \vdots\\
        \bm{W}_{\cdot,L_1}+ (f_{L_2}-\alpha_n)\bm{R}_{L_1+\delta,1}(1)+(f_{L_2}-\alpha_n)^2\bm{R}_{L_1+\delta,2}(1)+\ldots+(f_{L_2}-\alpha_n)^X\bm{R}_{L_1+\delta,X}(1)
    \end{bmatrix},
\end{align}
and
\begin{align}
    \bm{S}_n(2)=\begin{bmatrix}
        \bm{W}_{\cdot,L_1+1} + (f_1-\alpha_n)\bm{R}_{11}(2)+(f_1-\alpha_n)^2\bm{R}_{12}(2)+\ldots+(f_1-\alpha_n)^XR_{1X}(2)\\
        \bm{W}_{\cdot,L_1+2}+(f_2-\alpha_n)\bm{R}_{21}(2)+(f_2-\alpha_n)^2\bm{R}_{22}(2)+\ldots+(f_2-\alpha_n)^X\bm{R}_{2X}(2)\\
        \vdots\\
        \bm{W}_{\cdot,L_1+L_2}+ (f_{L_2}-\alpha_n)\bm{R}_{L_2,1}(2)+(f_{L_2}-\alpha_n)^2\bm{R}_{L_2,2}(2)+\ldots+(f_{L_2}-\alpha_n)^X\bm{R}_{L_2,X}(2)\\
    \end{bmatrix}.
\end{align}

Using the same procedure as in the previous section, the user sends the following query to the $n$th server
\begin{align}
    \bm{Q}_n^{[\theta]}=\begin{bmatrix}
        \frac{1}{f_1-\alpha_n}\left(\bm{e}_{\theta}+(f_1-\alpha_n)\bm{Z}_{11}+\ldots +(f_1-\alpha_n)^M \bm{Z}_{1M}\right)\\
        \vdots\\
        \frac{1}{f_{L_2}-\alpha_n}\left(\bm{e}_{\theta}+(f_{L_2}-\alpha_n)\bm{Z}_{L_21}+\ldots+ (f_{L_2}-\alpha_n)^M \bm{Z}_{L_2M}\right)
    \end{bmatrix}.
\end{align}
The transmission matrix is now slightly modified as follows
\begin{align}
    \mathtt{G}'(\bm{u},\bm{v}) = \begin{bmatrix}
        0_N& \mathtt{I}_N
    \end{bmatrix}\underbrace{\begin{bmatrix}
        \mathtt{I}_N&0&0\\
        0&\mathtt{V}_{2L_2+2X+2M-N}(\bm{b}_{[2L_2+2X+2M-N]})& 0 \\
        0&0& I_{2U}
    \end{bmatrix}
        \mathtt{M}^{-1}}_{\mathtt{M}_2^{-1}},
\end{align}
where $\mathtt{M}$ is the same as \eqref{main_transfer}.

The received symbols at the user side can be derived as follows
\begin{align}
    \bm{y} &= \begin{bmatrix}
        0_N& \mathtt{I}_N
    \end{bmatrix}\begin{bmatrix}
        \mathtt{I}_N&0&0\\
        0&\mathtt{V}_{2L_2+2X+2M-N}(\bm{b}_{[2L_2+2X+2M-N]})& 0 \\
        0&0& \mathtt{I}_{2U}
    \end{bmatrix}\begin{bmatrix}
        \bm{I}_1(1)\\
        \bm{I}_1(2)\\ \bm{r}^{[\theta]}_{[\delta]}\\\bm{W}_{\theta}(1)\\ \bm{W}_{\theta}(2)\\ \bm{I}_2(1)\\ \bm{I}_2(2)\\ \bm{\Delta}_U'(1)\\ \bm{\Delta}_U'(2)
    \end{bmatrix}\\
    &= \begin{bmatrix}
        \mathtt{V}_{2L_2+2X+2M-N}(\bm{b}_{[2L_2+2X+2M-N]})\begin{bmatrix}
            \bm{r}^{[\theta]}_{[\delta]}\\\bm{W}_{\theta}(1)\\ \bm{W}_{\theta}(2)\\ \bm{I}_2(1)\\ \bm{I}_2(2)
        \end{bmatrix}\\
        \bm{\Delta}_U'(1)\\
        \bm{\Delta}_U'(2)
    \end{bmatrix}.
\end{align}
Thus, since the first instance has $L_1$ useful symbols of the required message symbols and the second instance has $L_2$ useful symbols, the rate is given by $R_Q = \frac{L_1+L_2}{N} =  2\big(\frac{N-X-M-U-\frac{\delta}{2}}{N}\big)$.

\subsection{Regime 3: $U+E < X+M < \frac{N}{2}$}
In this regime, the user chooses $T_1 \geq T$ and $T_2 \geq T$ such that $T_1 +T_2+2X = N$ and $T_1 \leq T_2 \leq T_1+1$. Let $L_1 = \lceil\frac{N}{2}\rceil -U-E$ and $L_2 = \lfloor\frac{N}{2}\rfloor -U$. Thus, the storage is given by
\begin{align}
    \bm{S}_n(1) = \begin{bmatrix}
        \bm{r}_{\cdot,1}+ (f_1-\alpha_n)\bm{R}_{11}(1)+(f_{1}-\alpha_n)^2\bm{R}_{12}(1)+\ldots+(f_{1}-\alpha_n)^X\bm{R}_{1X}(1)\\
        \vdots\\
        \bm{r}_{\cdot,E}+ (f_{E}-\alpha_n)\bm{R}_{{E}1}(1)+(f_{E}-\alpha_n)^2\bm{R}_{{E}2}(1)+\ldots+(f_{E}-\alpha_n)^X\bm{R}_{{E}X}(1)\\
        \bm{W}_{\cdot,1} + (f_{E+1}-\alpha_n)\bm{R}_{E+1,1}(1)+(f_{E+1}-\alpha_n)^2\bm{R}_{E+1,2}(1)+\ldots+(f_{E+1}-\alpha_n)^X\bm{R}_{E+1,X}(1)\\
        \bm{W}_{\cdot,2} +(f_{E+2}-\alpha_n)\bm{R}_{E+2,1}(1)+(f_{E+2}-\alpha_n)^2\bm{R}_{E+2,2}(1)+\ldots+(f_{E+2}-\alpha_n)^X\bm{R}_{E+2,X}(1)\\
        \vdots\\
        \bm{W}_{\cdot,L_1}+ (f_{L_2}-\alpha_n)\bm{R}_{L_1+E,1}(1)+(f_{L_2}-\alpha_n)^2\bm{R}_{L_1+E,2}(1)+\ldots+(f_{L_2}-\alpha_n)^X\bm{R}_{L_1+E,X}(1)
    \end{bmatrix},
\end{align}
and
\begin{align}
    \bm{S}_n(2)=\begin{bmatrix}
        \bm{W}_{\cdot,L_1+1} + (f_1-\alpha_n)\bm{R}_{11}(2)+(f_1-\alpha_n)^2\bm{R}_{12}(2)+\ldots+(f_1-\alpha_n)^X\bm{R}_{1X}(2)\\
        \bm{W}_{\cdot,L_1+2}+(f_2-\alpha_n)\bm{R}_{21}(2)+(f_2-\alpha_n)^2\bm{R}_{22}(2)+\ldots+(f_2-\alpha_n)^X\bm{R}_{2X}(2)\\
        \vdots\\
        \bm{W}_{\cdot,L_1+L_2}+ (f_{L_2}-\alpha_n)\bm{R}_{L_2,1}(2)+(f_{L_2}-\alpha_n)^2\bm{R}_{L_2,2}(2)+\ldots+(f_{L_2}-\alpha_n)^X\bm{R}_{L_2,X}(2)\\
    \end{bmatrix}.
\end{align}

The queries are given by
\begin{align}
    \bm{Q}_n^{[\theta]}(1)=\begin{bmatrix}
        \frac{1}{f_1-\alpha_n}\left(\bm{e}_{\theta}+(f_1-\alpha_n)\bm{Z}_{11}(1)+\ldots +(f_1-\alpha_n)^{T_1} \bm{Z}_{1{T_1}}(1)\right)\\
        \vdots\\
        \frac{1}{f_{L_2}-\alpha_n}\left(\bm{e}_{\theta}+(f_{L_2}-\alpha_n)\bm{Z}_{{L_2}1}(1)+\ldots+ (f_{L_2}-\alpha_n)^{T_1} \bm{Z}_{{L_2}{T_1}}(1)\right)
    \end{bmatrix},
\end{align}
and 
\begin{align}
    \bm{Q}_n^{[\theta]}(2)=\begin{bmatrix}
        \frac{1}{f_1-\alpha_n}\left(\bm{e}_{\theta}+(f_1-\alpha_n)\bm{Z}_{11}(2)+\ldots +(f_1-\alpha_n)^{T_2} \bm{Z}_{1{T_2}}(2)\right)\\
        \vdots\\
        \frac{1}{f_{L_2}-\alpha_n}\left(\bm{e}_{\theta}+(f_{L_2}-\alpha_n)\bm{Z}_{{L_2}1}(2)+\ldots+ (f_{L_2}-\alpha_n)^{T_2} \bm{Z}_{{L_2}{T_2}}(2)\right)
    \end{bmatrix}.
\end{align}

Each answer instance from the $n$th server is given by
\begin{align}
    A_n^{[\theta]}(i)  = \bm{S}_n(i)^t \bm{Q}_n^{[\theta]}(i)+\sum_{j=1}^{X+T_i}\alpha_n^{i-1} Z'_j(i), \quad i \in [2].
\end{align}

The transmission matrix for this regime is the same as the previous section with $\delta=E$, and $2X+2M = N$, i.e.,
\begin{align}
    \mathtt{G}'(\bm{u},\bm{v}) = \begin{bmatrix}
        0_N& \mathtt{I}_N
    \end{bmatrix}\underbrace{\begin{bmatrix}
        \mathtt{I}_N&0&0\\
        0&\mathtt{V}_{L_1+L_2+E}(\bm{b}_{[L_1+L_2+E]})& 0 \\
        0&0& \mathtt{I}_{2U}
    \end{bmatrix}
        \mathtt{M}^{-1}}_{\mathtt{M}_2^{-1}}.
\end{align}

Thus, the received symbols at the user side can be derived as follows
\begin{align}
    \bm{y} &= \begin{bmatrix}
        0_N& \mathtt{I}_N
    \end{bmatrix}\begin{bmatrix}
        \mathtt{I}_N&0&0\\
        0&\mathtt{V}_{L_1+L_2+E}(\bm{b}_{[L_1+L_2+E]})& 0 \\
        0&0& \mathtt{I}_{2U}
    \end{bmatrix}\begin{bmatrix}
        \bm{I}_1(1)\\\bm{I}_1(2)\\ \bm{r}^{[\theta]}_{[E]}\\\bm{W}_{\theta}(1)\\ \bm{W}_{\theta}(2)\\ \bm{\Delta}_U'(1)\\ \bm{\Delta}_U'(2)
    \end{bmatrix}\\
    &= \begin{bmatrix}
        \mathtt{V}_{L_1+L_2+E}(\bm{b}_{[L_1+L_2+E]})\begin{bmatrix}
            \bm{r}^{[\theta]}_{[E]}\\\bm{W}_{\theta}(1)\\ \bm{W}_{\theta}(2)
        \end{bmatrix}\\
        \bm{\Delta}_U'(1)\\
        \bm{\Delta}_U'(2)
    \end{bmatrix}.
\end{align}
Then, by applying the same calculation steps, the rate is $R_Q = \frac{L_1+L_2}{N} = \frac{N-2U-E}{N}$.

\begin{remark}
    If $T_1 = T_2$, then it is sufficient to send one query instance only to the servers.
\end{remark}

\begin{remark}
    Note that $T_1 \geq E$ and $T_2 \geq E$, since the following inequalities hold
    \begin{align}
        X+M < \frac{N}{2} 
        & \Leftrightarrow X+M \leq \lfloor\frac{N}{2}\rfloor \\
        &\Leftrightarrow X+T \leq \lfloor\frac{N}{2}\rfloor, \quad X+E \leq \lfloor\frac{N}{2}\rfloor\\
        & \Leftrightarrow 2X+2E \leq 2 \lfloor\frac{N}{2}\rfloor\\
        & \Rightarrow 2E \leq T_1+T_2, \qquad \textrm{since }2X+T_1+T_2=N\\
        & \Leftrightarrow 2E \leq 2T_1 +1\\
        & \Rightarrow E \leq T_1 \\
        & \Rightarrow E \leq T_2.
    \end{align}
\end{remark}

\subsection{Regime 4: $X+M < \frac{N}{2}$ and $U+E> X+M$}
In this setting, the usage of the classical scheme is preferable compared to the quantum scheme since there are not enough dimensions for transmitting message symbols using the quantum scheme or the rate of the classical scheme is better than the quantum scheme. For the sake of completeness, we present the classical scheme here. The sub-packetization length is $L=N-X-M-U$ and the storage is as in \eqref{classical_storage} and queries by the user are as in \eqref{classical_queries}. In addition, the answers are computed as 
\begin{align}
    A_n^{[\theta]} = \bm{S}_n^t \bm{Q}_n^{[\theta]}+ \sum_{i=1}^{X+M}\alpha_n^{i-1}Z'_i, \quad n \in [N] \setminus \mathcal{U}. 
\end{align}
Thus, the answers can be written collectively as
\begin{align}
    \begin{bmatrix}
        A_{i_1}^{[\theta]}\\
        \vdots\\
        A_{i_{N-U}}^{[\theta]}
    \end{bmatrix} = \begin{bmatrix}
        \frac{1}{f_1-\alpha_{i_1}}\!&\!\ldots\!&\!\frac{1}{f_L-\alpha_{i_1}}\!&\!1\!&\!\alpha_{i_1}\!&\!\ldots\!&\!\alpha_{i_1}^{X+M-1}\\
        \!\frac{1}{f_1-\alpha_{i_2}}\!&\!\ldots\!&\!\frac{1}{f_L-\alpha_{i_2}}\!&\!1\!&\!\alpha_{i_2}\!&\!\ldots\!&\!\alpha_{i_2}^{X+M-1}\\
        \!\vdots\!&\!\vdots\!&\!\vdots\!&\!\vdots\!&\!\vdots\!&\!\vdots\!&\!\vdots\\
        \!\frac{1}{f_1-\alpha_{i_{N-U}}}\!&\!\ldots\!&\!\frac{1}{f_L-\alpha_{i_{N-U}}}\!&\!1\!&\!\alpha_{i_{N-U}}\!&\!\ldots\!&\!\alpha_{i_{N-U}}^{X+M-1}
    \end{bmatrix}\begin{bmatrix}
        W_{\theta,1}\\
        \vdots\\
        W_{\theta,L}\\
        I_1+Z'_1\\
        \vdots\\
        I_{X+M}+Z'_{X+M}
    \end{bmatrix}.
\end{align}

\section{Achievable Scheme for QXBEUTSPIR}\label{Byzantine_scheme}
In this section, we only analyze the actions of the Byzantine servers on the $N$-sum box as a black box, i.e., they use the same quantum operations the honest servers use but can manipulate the transmitted dits. To extend this to all arbitrary quantum operations, we use the fact that any quantum operation can be written as a linear combination of the Pauli operators. Thus, the achievable rates in this scheme will not change even if the Byzantine servers apply any quantum operations on their transmitted dits. The concrete analysis for this observation can be found in \cite{Byzantine_journal}.  

\subsection{Regime 1: $H+M+B \geq \frac{N}{2}$ and $ E \leq 2H+2M+2B -N$}
Let $L = N-H-M-3B-U$ and the storage at the $n$th database, $\bm{S}_n(i)$, $i \in [2]$,
\begin{align}
    \bm{S}_n(i)=\begin{bmatrix}
        \bm{W}_{\cdot,1}(i) + (f_1-\alpha_n)\bm{R}_{11}(i)+(f_1-\alpha_n)^2\bm{R}_{12}(i)+\ldots+(f_1-\alpha_n)^H\bm{R}_{1H}(i)\\
        \bm{W}_{\cdot,2}(i) + (f_2-\alpha_n)\bm{R}_{21}(i)+(f_2-\alpha_n)^2\bm{R}_{22}(i)+\ldots+(f_2-\alpha_n)^H\bm{R}_{2H}(i)\\
        \vdots\\
        \bm{W}_{\cdot,L}(i) + (f_L-\alpha_n)\bm{R}_{L1}(i)+(f_L-\alpha_n)^2\bm{R}_{L2}(i)+\ldots+(f_L-\alpha_n)^H\bm{R}_{LH}(i)\\
    \end{bmatrix}.
\end{align}
In addition, the noise used to provide symmetric privacy and protection against collaboration between the eavesdropper and the Byzantine servers is distributed to all servers as follows
\begin{align}
    \hat{Z}_n(i) = Z'_1(i) + \alpha_n Z'_2(i)+ \ldots + \alpha_n^{M+H-1} Z'_{H+M}(i)+ \alpha_n^{M+H}R'_{1}(i)+ \ldots+ \alpha_n^{M+H+B-1}R'_{B}(i),   
\end{align}
where $\{Z'_i,R'_j\}$ are uniform independent random variables. The queries are given by
\begin{align}
    \bm{Q}_n^{[\theta]}=\begin{bmatrix}
        \frac{1}{f_1-\alpha_n}\left(\bm{e}_{\theta}+(f_1-\alpha_n)\bm{Z}_{11}+\ldots +(f_1-\alpha_n)^M \bm{Z}_{1M}\right)\\
        \vdots\\
        \frac{1}{f_L-\alpha_n}\left(\bm{e}_{\theta}+(f_L-\alpha_n)\bm{Z}_{L1}+\ldots+ (f_L-\alpha_n)^M \bm{Z}_{LM}\right)
    \end{bmatrix},
\end{align}
where $\bm{e}_{\theta}$ is a vector of length $K$ with $1$ in the $\theta$th index and zero otherwise, and $Z_{ij}$ are uniform independent random vectors of length $K$ each, chosen by the user.

Let $\mathcal{B} = \{n_{i_1}, \ldots, n_{i_B}\}$ denote the indices of the Byzantine servers, and as in the previous section $\mathcal{U}$ denotes the indices of the unresponsive servers. Thus, the answer from the $n$th server, $n \in [N]\setminus \left(\mathcal{B} \cup \mathcal{U}\right)$, is given by
\begin{align}
    A_n^{[\theta]}(i) = \bm{S}_n(i)^t \bm{Q}_n^{[\theta]}+ \hat{Z}_n(i), 
\end{align}
and for $n \in \mathcal{B}$ is given by
\begin{align}
    A_n^{[\theta]}(i) = \eta_n(i) = \bm{S}_n(i)^t \bm{Q}_n^{[\theta]}+ \hat{Z}_n(i)+\Delta_n(i).
\end{align}
The answers from the $N-U$ servers, $i_1, \ldots, i_{N-U}$, for one instance, can be written collectively as shown in \cite{byzantine_1}, i.e., 
\begin{align}
    \begin{bmatrix}
        A_{i_1}^{[\theta]}\\
        \vdots\\
        A_{i_{N-U}}^{[\theta]}
    \end{bmatrix} &=\underbrace{\begin{bmatrix}
        \frac{1}{f_1-\alpha_{i_1}} &\ldots&\frac{1}{f_L-\alpha_{i_1}}&1&\alpha_{i_1}&\ldots&\alpha_{i_1}^{N-U-L-1}\\
        \frac{1}{f_1-\alpha_{i_2}}&\ldots&\frac{1}{f_L-\alpha_{i_2}}&1&\alpha_{i_2}&\ldots&\alpha_{i_2}^{N-U-L-1}\\
        \vdots&\vdots&\vdots&\vdots&\vdots&\vdots&\vdots\\
        \frac{1}{f_1-\alpha_{i_{N-U}}}&\ldots&\frac{1}{f_L-\alpha_{i_{N-U}}}&1&\alpha_{i_{N-U}}&\ldots&\alpha_{i_{N-U}}^{N-U-L-1}
    \end{bmatrix}}_{\mathtt{CSA}_{N-U, L,N-U}}\begin{bmatrix}
        \bm{W}_{\theta}\\
        \bm{I}_{[H+M]}+\bm{Z}'_{[H+M]}\\
        \bm{R}'_{[B]}\\
        \bm{0}_{2B}
    \end{bmatrix} + \bm{\Delta}_B \\
    &=\mathtt{CSA}_{N-U, L,N-U} \bm{x},
\end{align}
where 
\begin{align}
    \bm{x} =\begin{bmatrix}
        \bm{W}_{\theta}\\
        \bm{I}_{[H+M]}+\bm{Z}'_{[H+M]}\\
        \bm{R}'_{[B]}\\
        \bm{0}_{2B}
    \end{bmatrix}  + \mathtt{CSA}_{N-U, L,N-U}^{-1}\bm{\Delta}_B, 
\end{align}
and $\bm{\Delta}_B$ is a vector of length $N-U$ with only $B$ non-zero elements.

Based on the analysis in the previous section about the unresponsive servers, the two answer instances can be written as follows
\begin{align}\label{answers_collected}
\bm{A}^{[\theta]}=& \underbrace{\begin{bmatrix}
     \mathtt{\Lambda}_1(:,1:\lceil\frac{N}{2}\rceil)& 0&\mathtt{\Gamma}_1&0&\mathtt{\Lambda}_1(:,a_1 :H+M+3B) &0&\mathtt{\Omega}&0\\
    0&\mathtt{\Lambda}_2(:,1:\lfloor\frac{N}{2}\rfloor)&0&\mathtt{\Gamma}_2&0&\mathtt{\Lambda}_2(:,a_2 :H+M+3B)&0&\mathtt{\Omega}
    \end{bmatrix}}_{\mathtt{M}'} \nonumber\\
    &\begin{bmatrix}
        \Tilde{\bm{I}}_1(1)^t,\Tilde{\bm{I}}_1(2)^t, \Tilde{\bm{W}}_{\theta}(1)^t,\Tilde{\bm{W}}_{\theta}(2)^t, \Tilde{\bm{I}}_2(1)^t, \Tilde{\bm{I}}_2(2)^t, \Tilde{\bm{\Delta}}_B(1)^t,\Tilde{\bm{\Delta}}_B(2)^t,\bm{\Delta}_U'(1)^t, \bm{\Delta}_U'(2)^t
    \end{bmatrix}^t
\end{align}
where $\mathtt{\Lambda}_1 = \mathtt{GRS}_{N,H+M}^{\bm{\alpha}_{[N]},\bm{u}_{[N]}}$, $\mathtt{\Lambda}_2 = \mathtt{GRS}_{N,H+M}^{\bm{\alpha}_{[N]},\bm{v}_{[N]}}$, $\mathtt{\Gamma}_1 = \mathtt{GC}_{N,L}^{\bm{\alpha}_{[N]},\bm{u}_{[N]},\bm{f}_{[L]}}$, $\mathtt{\Gamma}_2 = \mathtt{GC}_{N,L}^{\bm{\alpha}_{[N]},\bm{v}_{[N]},\bm{f}_{[L]}}$, $a_1 = \lceil\frac{N}{2}\rceil+1$, $a_2 = \lfloor\frac{N}{2}\rfloor+1$, $\mathtt{\Omega} = [\bm{e}_{n_1}, \bm{e}_{n_2}, \ldots, \bm{e}_{n_U}]$, and 
\begin{align}
    \begin{bmatrix}
         \Tilde{\bm{W}}_{\theta}(i)\\\Tilde{\bm{I}}_1(i)\\ \Tilde{\bm{I}}_2(i)\\  \Tilde{\bm{\Delta}}_B(i) 
    \end{bmatrix}= \begin{bmatrix}
        \bm{W}_{\theta}(i)\\
        \hat{\bm{I}}_{[H+M+B]}(i)\\
        \bm{0}_{2B}
    \end{bmatrix} + \mathtt{CSA}_{N-U, L,N-U}^{-1}\bm{\Delta}_B(i), 
\end{align}
where
\begin{align}
    \hat{\bm{I}}_{[H+M+B]}(i) = \begin{bmatrix}
        \bm{I}_{[H+M]}+\bm{Z}'_{[H+M]}\\
        \bm{R}'_{[B]}
    \end{bmatrix}.
\end{align}

Since there are $E \leq 2H+2M+2B-N$ eavesdropped links, the servers can use a channel transmission matrix as follows
\begin{align}
    \mathtt{G}'(\bm{u},\bm{v}) = \begin{bmatrix}
        0_N& \mathtt{I}_N
    \end{bmatrix}\underbrace{\begin{bmatrix}
        \mathtt{I}_N&0&0\\
        0&\mathtt{V}_{2L+2H+2M+2B-N}(\bm{b}_{[2L+2H+2M+2B-N]})& 0 \\
        0&0& \mathtt{I}_{4B+2U}
    \end{bmatrix}
        \mathtt{M}^{'-1}}_{\mathtt{M}_1^{'-1}},
\end{align}
where $\mathtt{M}'$ is as shown in \eqref{answers_collected}. Now, the received symbols can be written as 
\begin{align}
    \bm{y} &= \begin{bmatrix}
        0_N& \mathtt{I}_N
    \end{bmatrix}\begin{bmatrix}
        \mathtt{I}_N&0&0\\
        0&\mathtt{V}_{2L+2H+2M+2B-N}(\bm{b}_{[2L+2H+2M+2B-N]})& 0 \\
        0&0& \mathtt{I}_{4B+2U}
    \end{bmatrix}\begin{bmatrix}
        \Tilde{\bm{I}}_1(1)\\
        \Tilde{\bm{I}}_1(2)\\
        \Tilde{\bm{W}}_{\theta}(1)\\
        \Tilde{\bm{W}}_{\theta}(2)\\
        \Tilde{\bm{I}}_2(1)\\
        \Tilde{\bm{I}}_2(2)\\ 
        \Tilde{\bm{\Delta}}_B(1)\\
        \Tilde{\bm{\Delta}}_B(2)\\
        \bm{\Delta}_U'(1)\\
        \bm{\Delta}_U'(2)
    \end{bmatrix}\\
    &=\begin{bmatrix}
    \mathtt{V}_{2L+2H+2M+2B-N}(\bm{b}_{[2L+2H+2M+2B-N]})\begin{bmatrix}
      \Tilde{\bm{W}}_{\theta}(1)\\
        \Tilde{\bm{W}}_{\theta}(2)\\
        \Tilde{\bm{I}}_2(1)\\
        \Tilde{\bm{I}}_2(2)  
    \end{bmatrix}\\
       \Tilde{\bm{\Delta}}_B(1)\\
        \Tilde{\bm{\Delta}}_B(2)\\
        \bm{\Delta}_U'(1)\\
        \bm{\Delta}_U'(2)
    \end{bmatrix} \label{last_eq_before_byzantine_dec}.
\end{align}

In Section~\ref{decoding_byzantine_section}, we show the decoding scheme that is applied on \eqref{last_eq_before_byzantine_dec} to decode the message symbols. Note that since $V_{2L+2H+2M+2B-N}(b_{[2L+2H+2M+2B-N]})$ is invertible, we have 
\begin{align}
    \bm{y}_1 & = \begin{bmatrix}
        \mathtt{V}_{2L+2H+2M+2B-N}(\bm{b}_{[2L+2H+2M+2B-N]})^{-1}&0\\
        0&\mathtt{I}_{4B+2U}
    \end{bmatrix}\bm{y} \\
    &=\begin{bmatrix} 
      \Tilde{\bm{W}}_{\theta}(1)\\
        \Tilde{\bm{W}}_{\theta}(2)\\
        \Tilde{\bm{I}}_2(1)\\
        \Tilde{\bm{I}}_2(2) \\
       \Tilde{\bm{\Delta}}_B(1)\\
        \Tilde{\bm{\Delta}}_B(2)\\
        \bm{\Delta}_U'(1)\\
        \bm{\Delta}_U'(2)
    \end{bmatrix}.
\end{align}

\subsection{Regime 2: $H+M+B \geq \frac{N}{2}$ and $E > 2H+2M+2B-N$}
In this case, we develop a scheme that can use the interference symbols along with the noise variables introduced in the storage to prevent the eavesdropper from decoding the messages. As the number of eavesdropped links is greater than the number of the received interference symbols, we need to introduce new noise variables to compensate for the difference as mentioned previously. Let $\delta = E -(2H+2M+2B-N)$, $L_1=N-H-M-3B-U-\delta$, and $L_2 =L_1+\delta$. Define $\bm{r}_{\cdot,1},\ldots,\bm{r}_{\cdot,\delta}$ as uniform random vectors. Then, the storage is given by 
\begin{align}
    \bm{S}_n(1) = \begin{bmatrix}
        \bm{r}_{\cdot,1}+ (f_1-\alpha_n)\bm{R}_{11}(1)+(f_{1}-\alpha_n)^2\bm{R}_{12}(1)+\ldots+(f_{1}-\alpha_n)^H\bm{R}_{1H}(1)\\
        \vdots\\
        \bm{r}_{\cdot,\delta}+ (f_{\delta}-\alpha_n)\bm{R}_{{\delta}1}(1)+(f_{\delta}-\alpha_n)^2\bm{R}_{{\delta}2}(1)+\ldots+(f_{\delta}-\alpha_n)^H\bm{R}_{{\delta}H}(1)\\
        \bm{W}_{\cdot,1} + (f_{\delta+1}-\alpha_n)\bm{R}_{\delta+1,1}(1)+(f_{\delta+1}-\alpha_n)^2\bm{R}_{\delta+1,2}(1)+\ldots+(f_{\delta+1}-\alpha_n)^H\bm{R}_{\delta+1,H}(1)\\
        \bm{W}_{\cdot,2} +(f_{\delta+2}-\alpha_n)\bm{R}_{\delta+2,1}(1)+(f_{\delta+2}-\alpha_n)^2\bm{R}_{\delta+2,2}(1)+\ldots+(f_{\delta+2}-\alpha_n)^H\bm{R}_{\delta+2,H}(1)\\
        \vdots\\
        \bm{W}_{\cdot,L_1}+ (f_{L_2}-\alpha_n)\bm{R}_{L_1+\delta,1}(1)+(f_{L_2}-\alpha_n)^2\bm{R}_{L_1+\delta,2}(1)+\ldots+(f_{L_2}-\alpha_n)^H\bm{R}_{L_1+\delta,H}(1)
    \end{bmatrix}
\end{align}
and
\begin{align}
    \bm{S}_n(2)=\begin{bmatrix}
        \bm{W}_{\cdot,L_1+1} + (f_1-\alpha_n)\bm{R}_{11}(2)+(f_1-\alpha_n)^2\bm{R}_{12}(2)+\ldots+(f_1-\alpha_n)^H\bm{R}_{1H}(2)\\
        \bm{W}_{\cdot,L_1+2}+(f_2-\alpha_n)\bm{R}_{21}(2)+(f_2-\alpha_n)^2\bm{R}_{22}+\ldots+(f_2-\alpha_n)^H\bm{R}_{2H}(2)\\
        \vdots\\
        \bm{W}_{\cdot,L_1+L_2}+ (f_{L_2}-\alpha_n)\bm{R}_{L_2,1}(2)+(f_{L_2}-\alpha_n)^2\bm{R}_{L_2,2}(2)+\ldots+(f_{L_2}-\alpha_n)^H\bm{R}_{L_2,H}(2)\\
    \end{bmatrix}.
\end{align}
Using the same procedure as in the previous section, the rate is given by $R_Q = \frac{L_1+L_2}{N} =  2\big(\frac{N-H-M-3B-U-\frac{\delta}{2}}{N}\big)$.

\subsection{Regime 3: $2B+U+E < H+M+B< \frac{N}{2}$}
In this regime, the user chooses $T_1 \geq T$ and $T_2 \geq T$ such that $T_1 +T_2+2H+2B = N$ and $T_1 \leq T_2 \leq T_1+1$. Let $L_1 = \lceil\frac{N}{2}\rceil - 3B-U-E$ and $L_2 = \lfloor\frac{N}{2}\rfloor - 3B-U$. Thus, the storage is given by
\begin{align}
    \bm{S}_n(1) = \begin{bmatrix}
        \bm{r}_{\cdot,1}+ (f_1-\alpha_n)\bm{R}_{11}(1)+(f_{1}-\alpha_n)^2\bm{R}_{12}(1)+\ldots+(f_{1}-\alpha_n)^H\bm{R}_{1H}(1)\\
        \vdots\\
        \bm{r}_{\cdot,E}+ (f_{E}-\alpha_n)\bm{R}_{{E}1}(1)+(f_{E}-\alpha_n)^2\bm{R}_{{E}2}(1)+\ldots+(f_{E}-\alpha_n)^H\bm{R}_{{E}H}(1)\\
        \bm{W}_{\cdot,1} + (f_{E+1}-\alpha_n)\bm{R}_{E+1,1}(1)+(f_{E+1}-\alpha_n)^2\bm{R}_{E+1,2}(1)+\ldots+(f_{E+1}-\alpha_n)^H\bm{R}_{E+1,H}(1)\\
        \bm{W}_{\cdot,2} +(f_{E+2}-\alpha_n)\bm{R}_{E+2,1}(1)+(f_{E+2}-\alpha_n)^2\bm{R}_{E+2,2}(1)+\ldots+(f_{E+2}-\alpha_n)^H\bm{R}_{E+2,H}(1)\\
        \vdots\\
        \bm{W}_{\cdot,L_1}+ (f_{L_2}-\alpha_n)\bm{R}_{L_1+E,1}(1)+(f_{L_2}-\alpha_n)^2\bm{R}_{L_1+E,2}(1)+\ldots+(f_{L_2}-\alpha_n)^H\bm{R}_{L_1+E,H}(1)
    \end{bmatrix},
\end{align}
and
\begin{align}
    \bm{S}_n(2)=\begin{bmatrix}
        \bm{W}_{\cdot,L_1+1} + (f_1-\alpha_n)\bm{R}_{11}(2)+(f_1-\alpha_n)^2\bm{R}_{12}(2)+\ldots+(f_1-\alpha_n)^H\bm{R}_{1H}(2)\\
        \bm{W}_{\cdot,L_1+2}+(f_2-\alpha_n)\bm{R}_{21}(2)+(f_2-\alpha_n)^2\bm{R}_{22}(2)+\ldots+(f_2-\alpha_n)^H\bm{R}_{2H}(2)\\
        \vdots\\
        \bm{W}_{\cdot,L_1+L_2}+ (f_{L_2}-\alpha_n)\bm{R}_{L_2,1}(2)+(f_{L_2}-\alpha_n)^2\bm{R}_{L_2,2}(2)+\ldots+(f_{L_2}-\alpha_n)^H\bm{R}_{L_2,H}(2)\\
    \end{bmatrix}.
\end{align}
In addition, we have the same interference symbols to provide symmetric privacy as follows
\begin{align}
    \hat{Z}_n(1) = Z'_1(1) + \alpha_n Z'_2(1)+ \ldots + \alpha_n^{T_1+H-1} Z'_{H+T_1}(1)+ \alpha_n^{T_1+H}R'_{1}(1)+ \ldots+ \alpha_n^{T_1+H+B-1}R'_{B}(1), \\
    \hat{Z}_n(2) = Z'_1(2) + \alpha_n Z'_2(2)+ \ldots + \alpha_n^{T_2+H-1} Z'_{H+T_2}(2)+ \alpha_n^{T_2+H}R'_{1}(2)+ \ldots+ \alpha_n^{T_2+H+B-1}R'_{B}(2)
\end{align}
The queries are given by
\begin{align}
    \bm{Q}_n^{[\theta]}(1)=\begin{bmatrix}
        \frac{1}{f_1-\alpha_n}\left(\bm{e}_{\theta}+(f_1-\alpha_n)\bm{Z}_{11}(1)+\ldots +(f_1-\alpha_n)^{T_1} \bm{Z}_{1{T_1}}(1)\right)\\
        \vdots\\
        \frac{1}{f_{L_2}-\alpha_n}\left(\bm{e}_{\theta}+(f_{L_2}-\alpha_n)\bm{Z}_{{L_2}1}(1)+\ldots+ (f_{L_2}-\alpha_n)^{T_1} \bm{Z}_{{L_2}{T_1}}(1)\right)
    \end{bmatrix},
\end{align}
and 
\begin{align}
    \bm{Q}_n^{[\theta]}(2)=\begin{bmatrix}
        \frac{1}{f_1-\alpha_n}\left(\bm{e}_{\theta}+(f_1-\alpha_n)\bm{Z}_{11}(2)+\ldots +(f_1-\alpha_n)^{T_2} \bm{Z}_{1{T_2}}(2)\right)\\
        \vdots\\
        \frac{1}{f_{L_2}-\alpha_n}\left(\bm{e}_{\theta}+(f_{L_2}-\alpha_n)\bm{Z}_{{L_2}1}(2)+\ldots+ (f_{L_2}-\alpha_n)^{T_2} \bm{Z}_{{L_2}{T_2}}(2)\right)
    \end{bmatrix}.
\end{align}
Each answer instance from the $n$th server is given by
\begin{align}
    A_n^{[\theta]}(i)  = \bm{S}_n(i)^t \bm{Q}_n^{[\theta]}(i)+\hat{Z}_n(i), \quad i \in [2].
\end{align}
Upon applying the same steps as the previous two sections, the rate becomes $R_Q = \frac{L_1+L_2}{N} = \frac{N-6B-2U-E}{N}$.

\subsection{Regime 4: $H+M+B< \frac{N}{2}$ and $2B+U+E> H+M+B$}
In this setting, the usage of the classical scheme is preferable since there are not enough dimensions for transmitting message symbols using the classical scheme or the rate provided by the classical scheme is higher than the quantum scheme. For the sake of completeness, we present the classical scheme here. The sub-packetization length is $L=N-H-M-U-3B$ and the storage can be written as in \eqref{classical_storage} and queries by the user as in \eqref{classical_queries}. In addition, the answers are computed as 
\begin{align}
    A_n^{[\theta]} = \bm{S}_n^t \bm{Q}_n^{[\theta]}+\hat{Z}_n(1), \quad n \in [N] \setminus \mathcal{B} \cup \mathcal{U}. 
\end{align}
Thus, the answers can be written collectively as
\begin{align}
    \begin{bmatrix}
        A_{i_1}^{[\theta]}\\
        \vdots\\
        A_{i_{N-U}}^{[\theta]}
    \end{bmatrix} = \begin{bmatrix}
        \frac{1}{f_1-\alpha_{i_1}}\!&\!\ldots\!&\!\frac{1}{f_L-\alpha_{i_1}}\!&\!1\!&\!\alpha_{i_1}\!&\!\ldots\!&\!\alpha_{i_1}^{H+M-1}\\
        \!\frac{1}{f_1-\alpha_{i_2}}\!&\!\ldots\!&\!\frac{1}{f_L-\alpha_{i_2}}\!&\!1\!&\!\alpha_{i_2}\!&\!\ldots\!&\!\alpha_{i_2}^{H+M-1}\\
        \!\vdots\!&\!\vdots\!&\!\vdots\!&\!\vdots\!&\!\vdots\!&\!\vdots\!&\!\vdots\\
        \!\frac{1}{f_1-\alpha_{i_{N-U}}}\!&\!\ldots\!&\!\frac{1}{f_L-\alpha_{i_{N-U}}}\!&\!1\!&\!\alpha_{i_{N-U}}\!&\!\ldots\!&\!\alpha_{i_{N-U}}^{H+M-1}
    \end{bmatrix}\bm{x},
\end{align}
where 
\begin{align}
    \bm{x} =\begin{bmatrix}
        \bm{W}_{\theta}\\
        \bm{I}_{[H+M]}+\bm{Z}'_{[H+M]}\\
        \bm{R}'_{[B]}\\
        \bm{0}_{2B}
    \end{bmatrix}  + \mathtt{CSA}_{N-U, L,N-U}^{-1}\bm{\Delta}_B. 
\end{align}

\begin{remark}
    Note that in this section we assumed that the Byzantine servers use Pauli $\mathsf{X}(\cdot)$, and $\mathsf{Z}(\cdot)$ operations when transmitting the answers. This can be extended in the same exact way that is used in \cite{Byzantine_journal}. To explain intuitively, Kraus operators of any quantum operation that output Hilbert space is the same as the input Hilbert space can be decomposed into a linear combination of Weyl operators $\mathtt{W}(\bm{v})$, and any quantum code that corrects errors with Kraus operators $\{E_i\}_i$ can also correct errors with $\{F_i\}$ where $F_i=\sum_jm_{ij}E_i$. For each term individually, the scheme has the same rate thus by using superposition the rate is the same as considering the Weyl operators only being applied by the Byzantine servers.
\end{remark}

\section{Decoding the Byzantine Errors}\label{decoding_byzantine_section}
In the previous sections, the schemes are analyzed such that the user retrieves the $N-U$ symbols after applying the measurements. However, as discussed earlier, those symbols are contaminated with Byzantine errors. In this section, the decoding technique after measurements is developed in order to perform error correction for the Byzantine errors. We will use the following definitions for ease of notation:
\begin{align}
    &\mathtt{CSA}_{N-U,L,N-U}^{-1} = [C_{i,j}],\\
   &\mathtt{D}(a,\mathcal{J}) = \begin{bmatrix}
        C_{1,\mathcal{J}(1)}& \ldots & C_{1,\mathcal{J}(B)}\\
        \vdots& \ddots &\vdots \\
        C_{a,\mathcal{J}(1)}& \ldots & C_{a,\mathcal{J}(B)}
    \end{bmatrix},\\
    &\mathtt{\Phi}(\mathcal{J}) =  \begin{bmatrix}
        C_{L+X+M+1,\mathcal{J}(1)}& \ldots & C_{L+X+M+1,\mathcal{J}(B)}\\
        \vdots& \ddots &\vdots \\
        C_{L+X+M+B,\mathcal{J}(1)}& \ldots & C_{L+X+M+B,\mathcal{J}(B)}
        \end{bmatrix},\\
    &\mathtt{\Psi}(\mathcal{J}) = \begin{bmatrix}
        C_{L+X+M+B+1,\mathcal{J}(1)}& \ldots & C_{L+X+M+B+1,\mathcal{J}(B)}\\
        \vdots& \ddots &\vdots \\
        C_{N-U,\mathcal{J}(1)}& \ldots & C_{N-U,\mathcal{J}(B)}
        \end{bmatrix},\\
    &\hat{\bm{\Delta}}_B = \begin{bmatrix}
        \Delta_1, \ldots, \Delta_B
    \end{bmatrix}^t.
\end{align}
Using these definitions, the following lemmas will be useful in the decoding procedure.

\begin{lemma}
    $\mathtt{\Psi}(\mathcal{J})$ is invertible for all $\mathcal{J} \subset [N]$, such that $|\mathcal{J}| = B$.
\end{lemma}

\begin{Proof}
    We prove this lemma by contradiction. Assume for any such $\mathcal{J}$ that $\mathtt{\Psi}(\mathcal{J})$ is not invertible. Thus, at least one of the columns can be written as a linear combination of the other columns, i.e., 
    \begin{align}
         \begin{bmatrix}
        C_{L+X+M+B+1,\mathcal{J}(m)}\\
        \vdots \\
        C_{N-U,\mathcal{J}(m)}
        \end{bmatrix} = \sum_{\substack{k=1\\k\neq m}}^B a_k          \begin{bmatrix}
        C_{L+X+M+B+1,\mathcal{J}(k)}\\
        \vdots \\
        C_{N-U,\mathcal{J}(k)}
        \end{bmatrix}. 
    \end{align}
    In addition, since $ \begin{bmatrix}
    C_{1,\mathcal{J}(1)},
    \ldots ,
    C_{N-U,\mathcal{J}(1)}
    \end{bmatrix}^t, \ldots, \begin{bmatrix} C_{1,\mathcal{J}(B)},
    \ldots ,
    C_{N-U,\mathcal{J}(B)}
    \end{bmatrix}^t $ are orthogonal to exactly $N-U-B$ rows of $\mathtt{CSA}_{N-U,L,N-U}$, we have
    \begin{align}
        \begin{bmatrix}
        \frac{1}{f_1-\alpha_{\mathcal{I}(1)}}\!&\!\ldots\!&\!\frac{1}{f_L-\alpha_{\mathcal{I}(1)}}\!&\!1\!&\!\alpha_{\mathcal{I}(1)}\!&\!\ldots\!&\!\alpha_{\mathcal{I}(1)}^{N-U-L-1}\\
        \!\frac{1}{f_1-\alpha_{\mathcal{I}(2)}}\!&\!\ldots\!&\!\frac{1}{f_L-\alpha_{\mathcal{I}(2)}}\!&\!1\!&\!\alpha_{\mathcal{I}(2)}\!&\!\ldots\!&\!\alpha_{\mathcal{I}(2)}^{N-U-L-1}\\
        \!\vdots\!&\!\vdots\!&\!\vdots\!&\!\vdots\!&\!\vdots\!&\!\vdots\!&\!\vdots\\
        \!\frac{1}{f_1-\alpha_{\mathcal{I}({N-U-B})}}\!&\!\ldots\!&\!\frac{1}{f_L-\alpha_{\mathcal{I}({N-U-B})}}\!&\!1\!&\!\alpha_{\mathcal{I}({N-U-B})}\!&\!\ldots\!&\!\alpha_{\mathcal{I}({N-U-B})}^{N-U-L-1}
    \end{bmatrix}\begin{bmatrix}
        C_{1,\mathcal{J}(1)}&\ldots & C_{1,\mathcal{J}(B)}\\
        \vdots&\vdots&\vdots\\
        C_{N-U,\mathcal{J}(1)}&\ldots & C_{N-U,\mathcal{J}(B)}
        \end{bmatrix}=0,
    \end{align}
    where $\mathcal{I} \cup \mathcal{J} = [N]$, and $\mathcal{I} \cap \mathcal{J} = \emptyset$. In addition, we have
    \begin{align}
    &\begin{bmatrix}
        \frac{1}{f_1-\alpha_{\mathcal{I}(1)}}\!&\!\ldots\!&\!\frac{1}{f_L-\alpha_{\mathcal{I}(1)}}\!&\!1\!&\!\ldots\!&\!\alpha_{\mathcal{I}(1)}^{X+M+B-1}\\
        \!\frac{1}{f_1-\alpha_{\mathcal{I}(2)}}\!&\!\ldots\!&\!\frac{1}{f_L-\alpha_{\mathcal{I}(2)}}\!&\!1\!&\!\ldots\!&\!\alpha_{\mathcal{I}(2)}^{X+M+B-1}\\
        \!\vdots\!&\!\vdots\!&\!\vdots\!&\!\vdots\!&\!\vdots\!&\!\vdots\\
        \!\frac{1}{f_1-\alpha_{\mathcal{I}({N-U-B})}}\!&\!\ldots\!&\!\frac{1}{f_L-\alpha_{\mathcal{I}({N-U-B})}}\!&\!1\!&\!\ldots\!&\!\alpha_{\mathcal{I}({N-U-B})}^{X+M+B-1}
    \end{bmatrix}\begin{bmatrix}
        C_{1,\mathcal{J}(1)}&\ldots & C_{1,\mathcal{J}(B)}\\
        \vdots&\vdots&\vdots\\
        C_{L+X+M+B,\mathcal{J}(1)}&\ldots & C_{L+X+M+B,\mathcal{J}(B)}
        \end{bmatrix}\nonumber\\ 
        &= -     \begin{bmatrix}
        \alpha_{\mathcal{I}(1)}^{X+M+B}\!&\!\ldots\!&\!\alpha_{\mathcal{I}(1)}^{N-U}\\
        \alpha_{\mathcal{I}(2)}^{X+M+B}\!&\!\ldots\!&\!\alpha_{\mathcal{I}(2)}^{N-U}\\
        \!\vdots\!&\!\vdots\!&\!\vdots\!\\
        \alpha_{\mathcal{I}({N-U-B})}^{X+M+B}\!&\!\ldots\!&\!\alpha_{\mathcal{I}({N-U-B})}^{N-U}
    \end{bmatrix}\begin{bmatrix}
        C_{L+X+M+B+1,\mathcal{J}(1)}&\ldots & C_{L+X+M+B+1,\mathcal{J}(B)}\\
        \vdots&\vdots&\vdots\\
        C_{N-U,\mathcal{J}(1)}&\ldots & C_{N-U,\mathcal{J}(B)}
        \end{bmatrix},
    \end{align}
    thus,
    \begin{align}
    &\begin{bmatrix}
        \frac{1}{f_1-\alpha_{\mathcal{I}(1)}}\!&\!\ldots\!&\!\frac{1}{f_L-\alpha_{\mathcal{I}(1)}}\!&\!1\!&\!\ldots\!&\!\alpha_{\mathcal{I}(1)}^{X+M+B-1}\\
        \!\frac{1}{f_1-\alpha_{\mathcal{I}(2)}}\!&\!\ldots\!&\!\frac{1}{f_L-\alpha_{\mathcal{I}(2)}}\!&\!1\!&\!\ldots\!&\!\alpha_{\mathcal{I}(2)}^{X+M+B-1}\\
        \!\vdots\!&\!\vdots\!&\!\vdots\!&\!\vdots\!&\!\vdots\!&\!\vdots\\
        \!\frac{1}{f_1-\alpha_{\mathcal{I}({N-U-B})}}\!&\!\ldots\!&\!\frac{1}{f_L-\alpha_{\mathcal{I}({N-U-B})}}\!&\!1\!&\!\ldots\!&\!\alpha_{\mathcal{I}({N-U-B})}^{X+M+B-1}
    \end{bmatrix}\begin{bmatrix}
        C_{1,\mathcal{J}(m)}\\
        \vdots\\
        C_{L+X+M+B,\mathcal{J}(m)}
        \end{bmatrix}\nonumber\\
    & = - \begin{bmatrix}
        \alpha_{\mathcal{I}(1)}^{X+M+B}\!&\!\ldots\!&\!\alpha_{\mathcal{I}(1)}^{N-U}\\
        \alpha_{\mathcal{I}(2)}^{X+M+B}\!&\!\ldots\!&\!\alpha_{\mathcal{I}(2)}^{N-U}\\
        \!\vdots\!&\!\vdots\!&\!\vdots\!\\
        \alpha_{\mathcal{I}({N-U-B})}^{X+M+B}\!&\!\ldots\!&\!\alpha_{\mathcal{I}({N-U-B})}^{N-U}
    \end{bmatrix}\begin{bmatrix}
        C_{L+X+M+B+1,\mathcal{J}(m)}\\
        \vdots\\
        C_{N-U,\mathcal{J}(m)}
        \end{bmatrix}\nonumber\\
        &=-\sum_{\substack{k=1\\k\neq m}}a_k \begin{bmatrix}
        \alpha_{\mathcal{I}(1)}^{X+M+B}\!&\!\ldots\!&\!\alpha_{\mathcal{I}(1)}^{N-U}\\
        \alpha_{\mathcal{I}(2)}^{X+M+B}\!&\!\ldots\!&\!\alpha_{\mathcal{I}(2)}^{N-U}\\
        \!\vdots\!&\!\vdots\!&\!\vdots\!\\
        \alpha_{\mathcal{I}({N-U-B})}^{X+M+B}\!&\!\ldots\!&\!\alpha_{\mathcal{I}({N-U-B})}^{N-U}
    \end{bmatrix}\begin{bmatrix}
        C_{L+X+M+B+1,\mathcal{J}(k)}\\
        \vdots\\
        C_{N-U,\mathcal{J}(k)}
        \end{bmatrix}\\
        &= \sum_{\substack{k=1\\k\neq m}}a_k\begin{bmatrix}
        \frac{1}{f_1-\alpha_{\mathcal{I}(1)}}\!&\!\ldots\!&\!\frac{1}{f_L-\alpha_{\mathcal{I}(1)}}\!&\!1\!&\!\ldots\!&\!\alpha_{\mathcal{I}(1)}^{X+M+B-1}\\
        \!\frac{1}{f_1-\alpha_{\mathcal{I}(2)}}\!&\!\ldots\!&\!\frac{1}{f_L-\alpha_{\mathcal{I}(2)}}\!&\!1\!&\!\ldots\!&\!\alpha_{\mathcal{I}(2)}^{X+M+B-1}\\
        \!\vdots\!&\!\vdots\!&\!\vdots\!&\!\vdots\!&\!\vdots\!&\!\vdots\\
        \!\frac{1}{f_1-\alpha_{\mathcal{I}({N-U-B})}}\!&\!\ldots\!&\!\frac{1}{f_L-\alpha_{\mathcal{I}({N-U-B})}}\!&\!1\!&\!\ldots\!&\!\alpha_{\mathcal{I}({N-U-B})}^{X+M+B-1}
    \end{bmatrix}\begin{bmatrix}
        C_{1,\mathcal{J}(k)}\\
        \vdots\\
        C_{L+X+M+B,\mathcal{J}(k)}
        \end{bmatrix}.
    \end{align}        
    This means that the first $L+X+M+B$ components of the $m$th column vector is as well a linear combination of the other columns with the same coefficients, i.e.,
    \begin{align}
    \begin{bmatrix}
        C_{1,\mathcal{J}(m)}\\
        \vdots\\
        C_{N-U,\mathcal{J}(m)}
        \end{bmatrix} = \sum_{\substack{k=1\\k \neq m}}a_k        
        \begin{bmatrix}
        C_{1,\mathcal{J}(k)}\\
        \vdots\\
        C_{N-U,\mathcal{J}(k)}
        \end{bmatrix}.
    \end{align}
    This is a contradiction since the inverse matrix of $\mathtt{CSA}_{N-U,L,N-U}$ is a full-rank matrix.
\end{Proof}

\begin{lemma}
    If $\mathtt{\Phi}(\mathcal{J}_1) \mathtt{\Psi}(\mathcal{J}_1)^{-1} \mathtt{\Psi}(\mathcal{J}_2)  \hat{\bm{\Delta}}_B = \mathtt{\Phi}(\mathcal{J}_2)\hat{\bm{\Delta}}_B$, then
    $\mathtt{D}(N-U-2B,\mathcal{J}_1)\mathtt{\Psi}(\mathcal{J}_1)^{-1} \mathtt{\Psi}(\mathcal{J}_2)  \hat{\bm{\Delta}}_B=   \mathtt{D}(N-U-2B,\mathcal{J}_2) \hat{\bm{\Delta}}_B $, and  $\mathtt{D}(N-U,\mathcal{J}_1)\mathtt{\Psi}(\mathcal{J}_1)^{-1} \mathtt{\Psi}(\mathcal{J}_2)  \hat{\bm{\Delta}}_B=   \mathtt{D}(N-U,\mathcal{J}_2) \hat{\bm{\Delta}}_B $.
\end{lemma}

\begin{Proof}
    First, for any $\mathcal{J}_1$, and $\mathcal{J}_2$, we have $|\mathcal{J}_1 \cup \mathcal{J}_2|\leq 2B$. Thus, there are at least $2B$ rows of $\mathtt{CSA}_{N-U,L,N-U}$ that are orthogonal to $C_{:,m}$, such that $ m \in \mathcal{J}_1 \cup \mathcal{J}_2$. Thus, as in the previous proof, we have
    \begin{align}
         &\begin{bmatrix}
        \frac{1}{f_1-\alpha_{\mathcal{I}(1)}}\!&\!\ldots\!&\!\frac{1}{f_L-\alpha_{\mathcal{I}(1)}}\!&\!1\!&\!\ldots\!&\!\alpha_{\mathcal{I}(1)}^{X+M-1}\\
        \!\frac{1}{f_1-\alpha_{\mathcal{I}(2)}}\!&\!\ldots\!&\!\frac{1}{f_L-\alpha_{\mathcal{I}(2)}}\!&\!1\!&\!\ldots\!&\!\alpha_{\mathcal{I}(2)}^{X+M-1}\\
        \!\vdots\!&\!\vdots\!&\!\vdots\!&\!\vdots\!&\!\vdots\!&\!\vdots\\
        \!\frac{1}{f_1-\alpha_{\mathcal{I}({N-U-B})}}\!&\!\ldots\!&\!\frac{1}{f_L-\alpha_{\mathcal{I}({N-U-2B})}}\!&\!1\!&\!\ldots\!&\!\alpha_{\mathcal{I}({N-U-2B})}^{X+M-1}
    \end{bmatrix}\begin{bmatrix}
        C_{1,\mathcal{J}(1)}&\ldots & C_{1,\mathcal{J}(B)}\\
        \vdots&\vdots&\vdots\\
        C_{L+X+M,\mathcal{J}(1)}&\ldots & C_{L+X+M,\mathcal{J}(B)}
        \end{bmatrix}\nonumber\\ 
        &= -     \begin{bmatrix}
        \alpha_{\mathcal{I}(1)}^{X+M}\!&\!\ldots\!&\!\alpha_{\mathcal{I}(1)}^{N-U}\\
        \alpha_{\mathcal{I}(2)}^{X+M}\!&\!\ldots\!&\!\alpha_{\mathcal{I}(2)}^{N-U}\\
        \!\vdots\!&\!\vdots\!&\!\vdots\!\\
        \alpha_{\mathcal{I}({N-U-2B})}^{X+M}\!&\!\ldots\!&\!\alpha_{\mathcal{I}({N-U-2B})}^{N-U}
    \end{bmatrix}\begin{bmatrix}
        C_{L+X+M+1,\mathcal{J}(1)}&\ldots & C_{L+X+M+1,\mathcal{J}(B)}\\
        \vdots&\vdots&\vdots\\
        C_{N-U,\mathcal{J}(1)}&\ldots & C_{N-U,\mathcal{J}(B)}
        \end{bmatrix},
    \end{align}
    which can be written as 
    \begin{align}
     &\begin{bmatrix}
        \frac{1}{f_1-\alpha_{\mathcal{I}(1)}}\!&\!\ldots\!&\!\frac{1}{f_L-\alpha_{\mathcal{I}(1)}}\!&\!1\!&\!\ldots\!&\!\alpha_{\mathcal{I}(1)}^{X+M-1}\\
        \!\frac{1}{f_1-\alpha_{\mathcal{I}(2)}}\!&\!\ldots\!&\!\frac{1}{f_L-\alpha_{\mathcal{I}(2)}}\!&\!1\!&\!\ldots\!&\!\alpha_{\mathcal{I}(2)}^{X+M-1}\\
        \!\vdots\!&\!\vdots\!&\!\vdots\!&\!\vdots\!&\!\vdots\!&\!\vdots\\
        \!\frac{1}{f_1-\alpha_{\mathcal{I}({N-U-B})}}\!&\!\ldots\!&\!\frac{1}{f_L-\alpha_{\mathcal{I}({N-U-2B})}}\!&\!1\!&\!\ldots\!&\!\alpha_{\mathcal{I}({N-U-2B})}^{X+M-1}
    \end{bmatrix}\mathtt{D}(N-U-2B,\mathcal{J}_1) \nonumber\\&= - \begin{bmatrix}
        \alpha_{\mathcal{I}(1)}^{X+M}\!&\!\ldots\!&\!\alpha_{\mathcal{I}(1)}^{N-U}\\
        \alpha_{\mathcal{I}(2)}^{X+M}\!&\!\ldots\!&\!\alpha_{\mathcal{I}(2)}^{N-U}\\
        \!\vdots\!&\!\vdots\!&\!\vdots\!\\
        \alpha_{\mathcal{I}({N-U-2B})}^{X+M}\!&\!\ldots\!&\!\alpha_{\mathcal{I}({N-U-2B})}^{N-U}
    \end{bmatrix}\begin{bmatrix}
        \mathtt{\Phi}(\mathcal{J}_1)\\
        \mathtt{\Psi}(\mathcal{J}_1)
    \end{bmatrix}.
    \end{align}
    By multiplying both sides by $\mathtt{\Psi}(\mathcal{J}_1)^{-1}\mathtt{\Psi}(\mathcal{J}_2)\hat{\bm{\Delta}}_B$, we have
    \begin{align}
     &\begin{bmatrix}
        \frac{1}{f_1-\alpha_{\mathcal{I}(1)}}\!&\!\ldots\!&\!\frac{1}{f_L-\alpha_{\mathcal{I}(1)}}\!&\!1\!&\!\ldots\!&\!\alpha_{\mathcal{I}(1)}^{X+M-1}\\
        \!\frac{1}{f_1-\alpha_{\mathcal{I}(2)}}\!&\!\ldots\!&\!\frac{1}{f_L-\alpha_{\mathcal{I}(2)}}\!&\!1\!&\!\ldots\!&\!\alpha_{\mathcal{I}(2)}^{X+M-1}\\
        \!\vdots\!&\!\vdots\!&\!\vdots\!&\!\vdots\!&\!\vdots\!&\!\vdots\\
        \!\frac{1}{f_1-\alpha_{\mathcal{I}({N-U-B})}}\!&\!\ldots\!&\!\frac{1}{f_L-\alpha_{\mathcal{I}({N-U-2B})}}\!&\!1\!&\!\ldots\!&\!\alpha_{\mathcal{I}({N-U-2B})}^{X+M-1}
    \end{bmatrix}\mathtt{D}(N-U-2B,\mathcal{J}_1)\mathtt{\Psi}(\mathcal{J}_1)^{-1}\mathtt{\Psi}(\mathcal{J}_2)\hat{\bm{\Delta}}_B \nonumber\\&= - \begin{bmatrix}
        \alpha_{\mathcal{I}(1)}^{X+M}\!&\!\ldots\!&\!\alpha_{\mathcal{I}(1)}^{N-U}\\
        \alpha_{\mathcal{I}(2)}^{X+M}\!&\!\ldots\!&\!\alpha_{\mathcal{I}(2)}^{N-U}\\
        \!\vdots\!&\!\vdots\!&\!\vdots\!\\
        \alpha_{\mathcal{I}({N-U-2B})}^{X+M}\!&\!\ldots\!&\!\alpha_{\mathcal{I}({N-U-2B})}^{N-U}
    \end{bmatrix}\begin{bmatrix}
        \mathtt{\Phi}(\mathcal{J}_1)\mathtt{\Psi}(\mathcal{J}_1)^{-1}\mathtt{\Psi}(\mathcal{J}_2)\hat{\bm{\Delta}}_B\\
        \mathtt{\Psi}(\mathcal{J}_2)\hat{\bm{\Delta}}_B
    \end{bmatrix}\\
    &=- \begin{bmatrix}
        \alpha_{\mathcal{I}(1)}^{X+M}\!&\!\ldots\!&\!\alpha_{\mathcal{I}(1)}^{N-U}\\
        \alpha_{\mathcal{I}(2)}^{X+M}\!&\!\ldots\!&\!\alpha_{\mathcal{I}(2)}^{N-U}\\
        \!\vdots\!&\!\vdots\!&\!\vdots\!\\
        \alpha_{\mathcal{I}({N-U-2B})}^{X+M}\!&\!\ldots\!&\!\alpha_{\mathcal{I}({N-U-2B})}^{N-U}
    \end{bmatrix}\begin{bmatrix}
        \mathtt{\Phi}(\mathcal{J}_2)\hat{\bm{\Delta}}_B\\
        \mathtt{\Psi}(\mathcal{J}_2)\hat{\bm{\Delta}}_B
    \end{bmatrix}\\
    &= \begin{bmatrix}
        \frac{1}{f_1-\alpha_{\mathcal{I}(1)}}\!&\!\ldots\!&\!\frac{1}{f_L-\alpha_{\mathcal{I}(1)}}\!&\!1\!&\!\ldots\!&\!\alpha_{\mathcal{I}(1)}^{X+M-1}\\
        \!\frac{1}{f_1-\alpha_{\mathcal{I}(2)}}\!&\!\ldots\!&\!\frac{1}{f_L-\alpha_{\mathcal{I}(2)}}\!&\!1\!&\!\ldots\!&\!\alpha_{\mathcal{I}(2)}^{X+M-1}\\
        \!\vdots\!&\!\vdots\!&\!\vdots\!&\!\vdots\!&\!\vdots\!&\!\vdots\\
        \!\frac{1}{f_1-\alpha_{\mathcal{I}({N-U-B})}}\!&\!\ldots\!&\!\frac{1}{f_L-\alpha_{\mathcal{I}({N-U-2B})}}\!&\!1\!&\!\ldots\!&\!\alpha_{\mathcal{I}({N-U-2B})}^{X+M-1}
    \end{bmatrix}\mathtt{D}(N-U-2B,\mathcal{J}_2)\hat{\bm{\Delta}}_B.
    \end{align}
    This concludes the first part of the lemma. The proof of the second part follows as
    \begin{align}
    \mathtt{D}(N-U,\mathcal{J}_2)\hat{\bm{\Delta}}_B &= \begin{bmatrix}
        \bm{V}(N-U-2B,\mathcal{J}_2)\\
        \mathtt{\Phi}(\mathcal{J}_1)\\
        \mathtt{\Psi}(\mathcal{J}_1)
    \end{bmatrix} \hat{\bm{\Delta}}_B\\
    &= \begin{bmatrix}
        \mathtt{D}(N-U-2B,\mathcal{J}_1)\mathtt{\Psi}(\mathcal{J}_1)^{-1}\mathtt{\Psi}(\mathcal{J}_2)\hat{\bm{\Delta}}_B\\
        \mathtt{\Phi}(\mathcal{J}_1)\mathtt{\Psi}(\mathcal{J}_1)^{-1}\mathtt{\Psi}(\mathcal{J}_2)\hat{\bm{\Delta}}_B\\
        \mathtt{\Psi}(\mathcal{J}_1)\mathtt{\Psi}(\mathcal{J}_1)^{-1}\mathtt{\Psi}(\mathcal{J}_2)\hat{\bm{\Delta}}_B
    \end{bmatrix}\\
    &= \begin{bmatrix}
        \mathtt{D}(N-U-2B,\mathcal{J}_1)\\
        \mathtt{\Phi}(\mathcal{J}_1)\\
        \mathtt{\Psi}(\mathcal{J}_1)
    \end{bmatrix}\mathtt{\Psi}(\mathcal{J}_1)^{-1}\mathtt{\Psi}(\mathcal{J}_2)\hat{\bm{\Delta}}_B\\
    &= \mathtt{D}(N-U,\mathcal{J}_1)\mathtt{\Psi}(\mathcal{J}_1)^{-1}\mathtt{\Psi}(\mathcal{J}_2)\hat{\bm{\Delta}}_B,
\end{align}
concluding the proof.
\end{Proof}

Now, assume that the real Byzantine servers are $\mathcal{J}_2$  and the user guess is $\mathcal{J}_1$. Since we have $2B$ symbols, we can do the following: First, the user estimates $\hat{\bm{\Delta}}_B$ using the first $B$ symbols and using the last $B$ symbols, the estimation is either confirmed or denied. Since $\mathtt{\Psi}(\mathcal{J}_1)$ is invertible, the user gets the following estimate $\bm{\overline{\Delta}}_B =\mathtt{\Psi}(\mathcal{J}_1)^{-1}\mathtt{\Psi}(\mathcal{J}_2)\hat{\bm{\Delta}}_B$. Afterwards, the user checks if $\mathtt{\Phi}(\mathcal{J}_1) \mathtt{\Psi}(\mathcal{J}_1)^{-1} \mathtt{\Psi}(\mathcal{J}_2)  \hat{\bm{\Delta}}_B = \mathtt{\Phi}(\mathcal{J}_2)\hat{\bm{\Delta}}_B$. If this is satisfied, then to remove the Byzantine effect from the $N-U$ servers, the user multiplies $\bm{\overline{\Delta}}_B$ by $\mathtt{D}(N-U,\mathcal{J}_1)$ and,
\begin{align}
    \begin{bmatrix}
         \bm{W}_{\theta}(i)\\
         \bm{I}_{[X+M-\gamma_i]}(i)+\bm{Z}'_{[X+M-\gamma_i]}(i)
    \end{bmatrix} = \bm{x}(1:L+X+M-\gamma_i) - (\mathtt{D}(N-U,\mathcal{J}_1)\bm{\overline{\Delta}}_B)(1:L+X+M-\gamma_i),
\end{align}
where $\gamma_1 = \lfloor\frac{N}{2}\rfloor$, and $\gamma_2 = \lceil\frac{N}{2}\rceil$.


\section{Proofs}\label{proofs}

\begin{lemma}
    The storage at the $n$th server given by
    \begin{align}
    \bm{S}_n=\begin{bmatrix}
        \bm{W}_{\cdot,1} + (f_1-\alpha_n)\bm{R}_{11}+(f_1-\alpha_n)^2\bm{R}_{12}+\ldots+(f_1-\alpha_n)^X\bm{R}_{1X}\\
        \bm{W}_{\cdot,2} + (f_2-\alpha_n)\bm{R}_{21}+(f_2-\alpha_n)^2\bm{R}_{22}+\ldots+(f_2-\alpha_n)^X\bm{R}_{2X}\\
        \vdots\\
        \bm{W}_{\cdot,L} + (f_L-\alpha_n)\bm{R}_{L1}+(f_L-\alpha_n)^2\bm{R}_{L2}+\ldots+(f_L-\alpha_n)^X\bm{R}_{LX}\\
    \end{bmatrix},
\end{align}
is secure against any $X$ communicating databases.
\end{lemma}

\begin{Proof}
    First, note that since each symbol and each message are encoded using different random variables, it suffices to prove security for only one symbol. Let the $\ell$th symbol of the $m$th message stored in the $n$th database be denoted by $\Tilde{W}_{m,\ell}(n)$, thus
    \begin{align}
        \Tilde{W}_{m,\ell}(n) = W_{m,\ell} + (f_{\ell}-\alpha_n)r_1+ \ldots + (f_{\ell}-\alpha_n)^Xr_X,
    \end{align}
    where $r_i$, $i \in [X]$ are i.i.d.~random variables generated uniformly at random. After sharing $\Tilde{W}_{m,\ell}(n_1), \ldots, \Tilde{W}_{m,\ell}(n_X)$, each server has
    \begin{align}
        \bm{W}_{\mathcal{X}} &= \begin{bmatrix}
             W_{m,\ell} + (f_{\ell}-\alpha_{n_1})r_1+ \ldots + (f_{\ell}-\alpha_{n_1})^Xr_X\\
             \vdots\\
             W_{m,\ell} + (f_{\ell}-\alpha_{n_X})r_1+ \ldots + (f_{\ell}-\alpha_{n_X})^Xr_X
        \end{bmatrix}\\
        &= \bm{1}W_{m,\ell} + \begin{bmatrix}
            B_1\\
            \vdots\\
            B_X
        \end{bmatrix} \underbrace{\begin{bmatrix}
            r_1\\
            \vdots\\
            r_X
        \end{bmatrix}}_{\bm{r}},
    \end{align}
    where $B_i = [f_{\ell}-\alpha_{n_i}, \ldots, (f_{\ell}-\alpha_{n_i})^X]$. Since each row $B_{\ell}$ is independent, we know that $I(B_{\ell}\bm{r};B_{\ell'}\bm{r}) = 0$, and $I(B_{\ell}\bm{r};B_{\ell'}\bm{r}| \{B_{k}\bm{r}\}_{k \in [X]\setminus\{\ell, \ell'\}}) = 0$ using Shannon's one-time pad theorem. Thus,
    \begin{align}
        I(W_{m,\ell}; \bm{W}_{\mathcal{X}}) &= \sum_{k \in [X]}I(W_{m,\ell}; W_{m,\ell}+ B_{k}\bm{r})\\
        &= \sum_{k \in [X]}I(W_{m,\ell}; B_{k}\bm{r}) = 0,
    \end{align}
    completing the proof.
\end{Proof}

\begin{lemma}\label{collusion_lemma}
    The queries transmitted to the $n$th server given by
    \begin{align}
    \bm{Q}_n^{[\theta]}=\begin{bmatrix}
        \frac{1}{f_1-\alpha_n}\left(\bm{e}_{\theta}+(f_1-\alpha_n)\bm{Z}_{11}+\ldots +(f_1-\alpha_n)^M \bm{Z}_{1M}\right)\\
        \vdots\\
        \frac{1}{f_L-\alpha_n}\left(\bm{e}_{\theta}+(f_L-\alpha_n)\bm{Z}_{L1}+\ldots+ (f_L-\alpha_n)^M \bm{Z}_{LM}\right)
    \end{bmatrix}=\begin{bmatrix}
        \bm{q}_{n,1}\\
        \vdots\\
        \bm{q}_{n,L}
    \end{bmatrix},
\end{align}
are private against any $T \leq M$ colluding servers.
\end{lemma}

\begin{Proof}
First, assume that $T=M$ which is the largest possible value for $T$. Now, since each $\bm{q}_{n,i}$, $i \in [L]$ is independent of all others (as the random vectors used at each query instance are independent) it is sufficient to consider $\bm{q}_{\mathcal{T},i} = [\bm{q}_{{n_1},i}, \ldots, \bm{q}_{{n_M},i}]$. This can be written as
\begin{align}
    \begin{bmatrix}
        \bm{q}_{{n_1},1}\\
        \vdots\\
        \bm{q}_{{n_M},1}
    \end{bmatrix} = \begin{bmatrix}
        \frac{1}{f_1 - \alpha_{n_1}}\\
        \vdots\\
        \frac{1}{f_1 - \alpha_{n_M}}
    \end{bmatrix}\bm{e}_{\theta} +\underbrace{ \begin{bmatrix}
        1& (f_1-\alpha_{n_1})& \ldots&(f_1-\alpha_{n_1})^{M-1}\\
        \vdots& \vdots& \ddots & \vdots\\
        1& (f_1-\alpha_{n_M})& \ldots&(f_1-\alpha_{n_M})^{M-1}
    \end{bmatrix}}_{\mathtt{B}}\underbrace{\begin{bmatrix}
        Z_{11}\\
        \vdots\\
        Z_{1M}
    \end{bmatrix}}_{\bm{Z}}.
\end{align}
As explained in the previous proof, the rows of $\mathtt{B}$ are independent from each other and with the same line of arguments, we reach the required conclusion. 
\end{Proof}

\begin{remark}
Note that Lemma~\ref{collusion_lemma} implies $I(\theta;\bm{Q}_{\mathcal{E}}^{[\theta]}) = 0$, for $|\mathcal{E}| \leq M$.    
\end{remark}

In the next part, we prove some properties related to the transmitted qudits and measurements. Although this can be proven using the Von Neumann entropy and quantum mutual information as in \cite{our_journal}, another direction is used here using the classical information theoretic quantities in order to provide readers with an alternative way to deal with quantum systems using the $N$-sum box abstraction. We prove the required properties for the scheme for Byzantine servers with dynamic eavesdroppers, since it is the general version. Recall that for the Byzantine scheme presented in Section~\ref{Byzantine_scheme}, the transmitted vector for the first regime is written as 
\begin{align}
    \bm{h}_1=\begin{bmatrix}
        \mathtt{I}_N&0&0\\
        0&\mathtt{V}_{2L+2H+2M-N}(\bm{b}_{[2L+2H+2M+2B-N]})& 0 \\
        0&0& \mathtt{I}_{4B+2U}
    \end{bmatrix} \underbrace{\begin{bmatrix}
        \Tilde{\bm{I}}_1(1)\\
        \Tilde{\bm{I}}_1(2)\\ 
        \Tilde{\bm{W}}_{\theta}(1)\\ 
        \Tilde{\bm{W}}_{\theta}(2)\\
        \Tilde{\bm{I}}_2(1)\\ 
        \Tilde{\bm{I}}_2(2)\\ 
        \Tilde{\bm{\Delta}}_B(1)\\ 
        \Tilde{\bm{\Delta}}_B(2)\\ 
        \bm{\Delta}_U'(1)\\ 
        \bm{\Delta}_U'(2)
    \end{bmatrix}}_{\bm{x}},
\end{align}
where 
\begin{align}
    \begin{bmatrix}
         \Tilde{\bm{W}}_{\theta}(i)\\
         \Tilde{\bm{I}}_1(i)\\
         \Tilde{\bm{I}}_2(i)\\
         \Tilde{\bm{\Delta}}_B(i) 
    \end{bmatrix}= \begin{bmatrix}
        \bm{W}_{\theta}(i)\\
        \hat{\bm{I}}_{[H+M+B]}(i)\\
        \bm{0}_{2B}
    \end{bmatrix} + \mathtt{CSA}_{N-U, L,N-U}^{-1}\bm{\Delta}_B(i), 
\end{align}
and
\begin{align}\label{r_1_comment}
    \hat{\bm{I}}_{[H+M+B]}(i) = \begin{bmatrix}
        \bm{I}_{[H+M]}(i)+\bm{Z}'_{[H+M]}(i)\\
        \bm{R}'_{[B]}(i)
    \end{bmatrix}.
\end{align}

Upon measurement, the user retrieves $N$ symbols of $\bm{h}_1$, counting the erasure errors as well. In this paper, we examine the vector $\bm{h}_1$ and show that symmetric privacy and eavesdropper security are maintained for whatever $N-U$ or $E \leq N-U$ linear combination that can come out of any measurements. 

\begin{lemma}
    The noise symbols used in masking are secure against any $B$
    Byzantine servers.
\end{lemma}

\begin{Proof}
    Let $\mathcal{B}=\{i_1,\ldots, i_B\}$ denote the indices of the Byzantine servers. If the $B$ Byzantine servers are cooperating, they will have the noise that should have been added to the generated answers if they were honest as
    \begin{align}
        \begin{bmatrix}
            \hat{Z}_{i_1} \\ \vdots \\ \hat{Z}_{i_B}
        \end{bmatrix}&=\begin{bmatrix}
            1 & \ldots & \alpha_{i_1}^{H+M-1} &\alpha_{i_1}^{H+M} & \ldots & \alpha_{i_1}^{H+M+B-1} \\
            \vdots & \ddots & \vdots &\vdots & \ddots & \vdots \\
            1 & \ldots & \alpha_{i_B}^{H+M-1} &\alpha_{i_B}^{H+M} & \ldots & \alpha_{i_B}^{H+M+B-1} 
        \end{bmatrix}\begin{bmatrix}
            Z'_1 \\ \vdots \\ Z'_{H+M} \\ R'_1 \\ \vdots \\ R'_{B}
        \end{bmatrix} \\
        &=\begin{bmatrix}
            1 & \ldots & \alpha_{i_1}^{H+M-1} \\
            \vdots & \ddots & \vdots \\
            1 & \ldots & \alpha_{i_B}^{H+M-1}  
        \end{bmatrix}\begin{bmatrix}
            Z'_1 \\ \vdots \\ Z'_{H+M} 
        \end{bmatrix}+\begin{bmatrix}
            \alpha_{i_1}^{H+M} & \ldots & \alpha_{i_1}^{H+M+B-1} \\
           \vdots & \ddots & \vdots \\
            \alpha_{i_B}^{H+M} & \ldots & \alpha_{i_B}^{H+M+B-1} 
        \end{bmatrix}\begin{bmatrix}
            R'_1 \\ \vdots \\ R'_{B}
        \end{bmatrix}, \label{byzantinesymnoise}
    \end{align}
    for each instance of the answers, however, as $Z'_1(1),\ldots,Z'_{H+M}(1),R'_1(1),\ldots,R'_B(1)$ are independent of $Z'_1(2),\ldots,Z'_{H+M}(2),R'_1(2),\ldots,R'_B(2)$, it suffices to focus on only one instance to show the required result, as the other instance will follow from the same arguments.

    For brevity, let
    \begin{align}
         \bm{\alpha}_{\mathcal{B}}=\begin{bmatrix}
            \alpha_{i_1} \\ \vdots \\ \alpha_{i_B}
        \end{bmatrix}.
    \end{align}
    Moreover, let $\mathtt{V}(\bm{\alpha}_{\mathcal{B}})_{B \times H+M}$ denote the Vandermonde matrix of ${B \times H+M}$ dimensions and $\bm{\alpha}_{\mathcal{B}}$ as the interpolation points. Thus, \eqref{byzantinesymnoise} in this notation is written as
    \begin{align}
        \hat{\bm{Z}}_{\mathcal{B}}=\mathtt{V}(\bm{\alpha}_{\mathcal{B}})_{B \times H+M}\bm{Z'}_{[H+M]}+\diag(\bm{\alpha}_{\mathcal{B}})^{H+M}\mathtt{V}(\bm{\alpha}_{\mathcal{B}})_{B \times B}\bm{R}'_{[B]},
    \end{align}
    which means that
    \begin{align}
         \mathbb{P}(\hat{\bm{Z}}_{\mathcal{B}}=\hat{\bm{z}})&=\mathbb{P}(\mathtt{V}(\bm{\alpha}_{\mathcal{B}})_{B \times H+M}\bm{Z}'_{[H+M]}+\diag(\bm{\alpha}_{\mathcal{B}})^{H+M}\mathtt{V}(\bm{\alpha}_{\mathcal{B}})_{B \times B}\bm{R'}_{[B]}=\hat{\bm{z}})\\
         &=\frac{1}{q^B}\frac{1}{q^{H+M}}q^{H+M}=\frac{1}{q^B},
    \end{align}
    where the last line follows from the fact that, for any value of $\bm{R}'_{[B]}$, there exists a $\bm{Z}'_{[H+M]}$ satisfying the event inside the probability. This showcases that $\hat{\bm{Z}}_{\mathcal{B}}$ is uniformly distributed.

Now, also note that
\begin{align}
    \mathbb{P}(&\hat{\bm{Z}}_{\mathcal{B}}
    =\hat{\bm{z}},\bm{Z}'_{[H+M]}=\bm{z}') \nonumber \\
    &=\mathbb{P}(\mathtt{V}(\bm{\alpha}_{\mathcal{B}})_{B \times H+M}\bm{Z}'_{[H+M]}+\diag(\bm{\alpha}_{\mathcal{B}})^{H+M}\mathtt{V}(\bm{\alpha}_{\mathcal{B}})_{B \times B}\bm{R}'_{[B]}=\hat{\bm{z}},\bm{Z}'_{[H+M]}=\bm{z}') \\
    &=\mathbb{P}(\diag(\bm{\alpha}_{\mathcal{B}})^{H+M}\mathtt{V}(\bm{\alpha}_{\mathcal{B}})_{B \times B}\bm{R}'_{[B]}=\hat{\bm{z}}-\mathtt{V}(\bm{\alpha}_{\mathcal{B}})_{B \times H+M}\bm{z}',\bm{Z}'_{[H+M]}=\bm{z}') \\
    &=\mathbb{P}(\bm{R}'_{[B]}=(\diag(\bm{\alpha}_{\mathcal{B}})^{H+M}\mathtt{V}(\bm{\alpha}_{\mathcal{B}})_{B \times B})^{-1}\left(\hat{\bm{z}}-\mathtt{V}(\bm{\alpha}_{\mathcal{B}})_{B \times H+M}\bm{z}'\right),\bm{Z}'_{[H+M]}=\bm{z}') \\
    &=\mathbb{P}\left(\bm{R}'_{[B]}=(\diag(\bm{\alpha}_{\mathcal{B}})^{H+M}\mathtt{V}(\bm{\alpha}_{\mathcal{B}})_{B \times B})^{-1}\left(\hat{\bm{z}}-\mathtt{V}(\bm{\alpha}_{\mathcal{B}})_{B \times H+M}\bm{z}'\right)\right)P(\bm{Z}'_{[H+M]}=\bm{z}')\\
    &=\frac{1}{q^B}\frac{1}{q^{H+M}} \\
    &=\mathbb{P}(\hat{\bm{Z}}=\hat{\bm{z}})P(\bm{Z}'_{[H+M]}=\bm{z}'),
    \end{align}
which shows that $\hat{\bm{Z}}$ is independent from $\bm{Z}'_{[H+M]}$. Moreover, by construction $Z'_1,\ldots,Z'_{H+M}$ are independent from the queries as well. Thus,
\begin{align}
    I(\hat{Z}_{i_1},\ldots,\hat{Z}_{i_B},\bm{Q}^{[\theta]}_{i_1},\ldots,\bm{Q}^{[\theta]}_{i_B};Z'_1,\ldots,Z'_{H+M})=0,
\end{align}
concluding the proof.
\end{Proof}

\begin{lemma}\label{lemma_masking_variables}
   The noise symbols used in protecting the {interference symbols} are secure against the user for any action performed by the $B$ Byzantine servers.
\end{lemma}

\begin{Proof}
    First, note that $B\leq \frac{N}{3}$. Let $\mathcal{L}_1,\mathcal{L}_2 \subset [H+M]$ with $|\mathcal{L}_1|=|\mathcal{L}_2|=\max\{0,H+M-\lfloor\frac{N}{2}\rfloor\}$ and $\mathcal{H}_1 =\mathcal{H}_2= \{\max\{1,\lfloor\frac{N}{2}\rfloor - H-M+1\}, \ldots, B\} $. Then, we proceed as follows \begin{align}
    \frac{1}{2}&I(\bm{Z}'_{\mathcal{L}_1}(1),\bm{Z}'_{\mathcal{L}_2}(2);\bm{R}'_{\mathcal{H}_1}(1), \bm{R}'_{\mathcal{H}_2}(2), \hat{\bm{Z}}_{\mathcal{B}}(1),\hat{\bm{Z}}_{\mathcal{B}}(2))\nonumber\\
    &=I(\bm{Z}'_{\mathcal{L}_1}(1);\bm{R}'_{\mathcal{H}_1}(1), \hat{\bm{Z}}_{\mathcal{B}}(1))\\
    &=\underbrace{I(\bm{R}'_{\mathcal{H}_1}(1); \bm{Z}'_{\mathcal{L}_1}(1))}_{(a)} +\mathbbm{1}\big(H+M> \lfloor\frac{N}{2}\rfloor\big) I(\bm{Z}'_{\mathcal{L}_1}(1); \hat{\bm{Z}}_{\mathcal{B}}(1)|\bm{R}'_{\mathcal{H}_1}(1))\\
    &=\mathbbm{1}\big(H+M> \lfloor\frac{N}{2}\rfloor\big)I(\bm{Z}'_{\mathcal{L}_1}(1); \begin{bmatrix}
        1&\alpha_{n_1}&\ldots&\alpha_{n_1}^{H+M-1}\\
        \vdots&\vdots&\ddots&\vdots\\
        1&\alpha_{n_B}&\ldots&\alpha_{n_B}^{H+M-1}\end{bmatrix}\begin{bmatrix}
            Z'_1\\
            \vdots\\
            Z'_{H+M}
        \end{bmatrix})\\
        &=0 \label{last}
    \end{align}
    where $(a) = 0$ since these terms are generated uniformly at random and independent from each other, and \eqref{last} follows from the fact that $B < \frac{N}{2}$.
\end{Proof}

\begin{remark}
   { To make it easier to understand intuitively the proof of Lemma \ref{lemma_masking_variables}: $\mathcal{L}_1$ and $\mathcal{L}_2$ are the interference terms that are not dropped over the air, i.e., decoded by the user upon measurement. In addition, $\mathcal{H}_1$ and $\mathcal{H}_2$ are the remaining masking terms for the shared randomness between the servers. They are all contaminated by the Byzantine noise, however, since the user decodes the Byzantine noise as explained in Section \ref{decoding_byzantine_section}, we drop the noise effect. As a numerical example, if $N=17$, $X=5$, $T=4$, $B=2$, and $E=U=0$, then $\mathcal{L}_1 = \emptyset$, and $\mathcal{L}_2 = \{9\}$, i.e., no interference terms are received from the first instance while one interference term from the second instance is decoded by the user, $I_9(2)$. On the other hand $\mathcal{H}_1 = \mathcal{H}_2 = \{1,2\}$, thus $R_1(i)$, $R_2(i)$, for $i \in [2]$ are received.}
\end{remark}

\begin{lemma}
   For the first regime in Theorem~\ref{byzantine_eaves_thm}, symmetric privacy is guaranteed.
\end{lemma}

\begin{Proof}
    We proceed as follows,
    \begin{align}
    I(&\bm{W}_{[K]\setminus\theta}; (\mathtt{M}'_1\bm{x})_{[N+1,2N-2U]}) \nonumber\\
    &=I(\bm{W}_{[K]\setminus\theta}; (\bm{x})_{[N+1,2N-2U]}) \label{remark_rev_1}\\
    &= I(\bm{W}_{[K]\setminus\theta}; \Tilde{\bm{\Delta}}_B(1), \Tilde{\bm{\Delta}}_B(2))+ I(\bm{W}_{[K]\setminus\theta}; (\bm{x})_{[N+1,2N-2U-4B]}|\Tilde{\bm{\Delta}}_B(1), \Tilde{\bm{\Delta}}_B(2))\\
    & = \underbrace{I(\bm{W}_{[K]\setminus\theta}; \delta_1(1), \ldots, \delta_B(2))}_{(a)}+ I(\bm{W}_{[K]\setminus\theta}; (\bm{x})_{[N+1,2N-2U-4B]}|\delta_1(1), \ldots, \delta_B(2)) \label{comment_1_1}\\
    & =  I(\bm{W}_{[K]\setminus\theta}; \bm{W}_{\theta}(1),\bm{W}_{\theta}(2), \hat{\bm{I}}_2(1),\hat{\bm{I}}_2(2)|\delta_1(1), \ldots, \delta_B(2))\\
    &= I(\bm{W}_{[K]\setminus\theta}; \hat{\bm{I}}_2(1),\hat{\bm{I}}_2(2)|\delta_1(1), \ldots, \delta_B(2),\bm{W}_{\theta}(1),\bm{W}_{\theta}(2))\\
    &=I(\bm{W}_{[K]\setminus\theta}; \bm{Z'}_{\mathcal{L}_1}(1),\bm{Z}'_{\mathcal{L}_2}(2), \bm{R}'_{\mathcal{H}_1}(1),\bm{R}'_{\mathcal{H}_2}(2)|\delta_1(1), \ldots, \delta_B(2),\bm{W}_{\theta}(1),\bm{W}_{\theta}(2)) \label{r_1_1_comment}\\
    &=0\label{last_line},
    \end{align}
    where \eqref{comment_1_1} follows from the data processing inequality and the fact that the eavesdropper can also apply the same algorithm applied by the user to know the location of the Byzantine servers and their transmitted noise symbols,  $(a) = 0$ since all massages are secure against any $B$ servers, { \eqref{r_1_1_comment} is due to \eqref{r_1_comment} and the independence of the noise variables,} and \eqref{last_line} is due to independence between the masking variables and the messages and due to Lemma~\ref{lemma_masking_variables}. 
\end{Proof}

{
\begin{remark}
   To see why it is enough to consider $(\mathtt{M}'_ 1\bm{x})_{[N+1,2N-2U]}$ for the answers: Considering that the servers share the quantum state and encoding operation done by them, the answers the user gets will be elements of an orthonormal basis for the whole quantum system of all servers. This orthonormality then translates to perfect distinguishability in the measurements, which means that the measurement results completely describe the quantum answers the user has received. Thus, $(\mathtt{M}'_ 1\bm{x})_{[N+1,2N-2U]}$, which is the classical measurement result, is actually a description of the quantum answer the user has received.
\end{remark}

}

\begin{lemma}
    For the first regime in Theorem~\ref{byzantine_eaves_thm}, the scheme is secure against the dynamic eavesdropper.
\end{lemma}

\begin{Proof}
    First, recall that, in the first regime, the number of eavesdropped links is $E \leq 2M+2H +2B - N$. In addition, for the vector $\bm{x}$, we have at least $2H+2M+2B$ independent random variables, i.e., $\bm{Z}'_{[H+M]}$, and $\bm{R}'_{[B]}$. In addition, we have
    \begin{align}\label{helpful_fact_1}
        I(\bm{Z}'_{\mathcal{E}}(1); \hat{\bm{Z}}_{\mathcal{B}}(1)) = 0,
    \end{align}
    where $|\mathcal{E}| \leq E$ and $H+M> B$.
    
    Furthermore, $\mathtt{M}'_1 \bm{x}$ hides all the message symbols with $2M+2H+2B-N$ independent random variables from the interference symbols. Thus,
    \begin{align}
        I(\bm{W}_{[K]}, & \theta; (\mathtt{M}'_1\bm{x})_{\mathcal{E}_1},\bm{Q}_{\mathcal{E}_2}) \nonumber\\
        =& \underbrace{I(\theta;\bm{Q}_{\mathcal{E}_2})}_{ =0  \text{ (by Lemma~\ref{collusion_lemma})}}+ I( \theta; (\mathtt{M}'_1\bm{x})_{\mathcal{E}_1}|\bm{Q}_{\mathcal{E}_2})+\underbrace{I(\bm{W}_{[K]}; \bm{Q}_{\mathcal{E}_2}|\theta)}_{=0 \text{ (by assumption)}}+I(\bm{W}_{[K]}; (\mathtt{M}'_1\bm{x})_{\mathcal{E}_1}|\bm{Q}_{\mathcal{E}_2},\theta)\\
        =& I(\theta;\{Z'_{i},(\mathtt{V}\bm{y})_{j},\Delta_{k}(1), \Delta_{\ell}(2)\}_{i \in \mathcal{I},j \in \mathcal{J},k \in \mathcal{K},\ell \in \mathcal{L} \atop |\mathcal{I}\cup\mathcal{J}\cup \mathcal{K} \cup \mathcal{L}| = |\mathcal{E}_1|}|\bm{Q}_{\mathcal{E}_2})\nonumber\\
        &+I(\bm{W}_{[K]}; \{Z'_{i},(\mathtt{V}\bm{y})_{j},\Delta_{k}(1), \Delta_{\ell}(2)\}_{i \in \mathcal{I},j \in \mathcal{J},k \in \mathcal{K},\ell \in \mathcal{L} \atop |\mathcal{I}\cup\mathcal{J}\cup \mathcal{K} \cup \mathcal{L}| = |\mathcal{E}_1|}|\bm{Q}_{\mathcal{E}_2},\theta)\\
        =&I(\theta;\{\Delta_{k}(1), \Delta_{\ell}(2)\}_{k \in \mathcal{K},\ell \in \mathcal{L} \atop |\mathcal{K} \cup \mathcal{L}| \leq \min\{4B, |\mathcal{E}_1 \setminus\mathcal{I}\cup \mathcal{J}|\}}|\bm{Q}_{\mathcal{E}_2}) \nonumber\\ &+ I(\theta;\{Z'_{i},(\mathtt{V}\bm{y})_{j}\}_{i \in \mathcal{I},j \in \mathcal{J} \atop |\mathcal{I}\cup\mathcal{J}| \leq  |\mathcal{E}_1|}|\bm{Q}_{\mathcal{E}_2},\{\Delta_{k}(1), \Delta_{\ell}(2)\}_{k \in \mathcal{K},\ell \in \mathcal{L} \atop |\mathcal{K} \cup \mathcal{L}| \leq \min\{4B, |\mathcal{E}_1 \setminus \mathcal{I} \cup \mathcal{J}|\}})\nonumber\\
        &+I(\bm{W}_{[K]}; \{\Delta_{k}(1), \Delta_{\ell}(2)\}_{k \in \mathcal{K},\ell \in \mathcal{L} \atop |\mathcal{K} \cup \mathcal{L}| \leq \min\{4B, |\mathcal{E}_1 \setminus\mathcal{I}\cup \mathcal{J}|\}}|\bm{Q}_{\mathcal{E}_2},\theta)\label{comment_1}\nonumber\\
        &+I(\bm{W}_{[K]};\{Z'_{i},(\mathtt{V}\bm{y})_{j}\}_{i \in \mathcal{I},j \in \mathcal{J} \atop |\mathcal{I}\cup\mathcal{J}| \leq  |\mathcal{E}_1|}|\bm{Q}_{\mathcal{E}_2},\theta,\{\Delta_{k}(1), \Delta_{\ell}(2)\}_{k \in \mathcal{K},\ell \in \mathcal{L} \atop |\mathcal{K} \cup \mathcal{L}| \leq \min\{4B, |\mathcal{E}_1 \setminus \mathcal{I} \cup \mathcal{J}|\}})\\
         \leq &I(\theta, \bm{Q}_{\mathcal{B}}|\bm{Q}_{\mathcal{E}_2})+I(\bm{W}_{[K]}, \bm{S}_{\mathcal{B}}|\bm{Q}_{\mathcal{E}_2},\theta)=0 \label{comment_2},
    \end{align}
    where $\mathcal{I} \subset [N]$, $\mathcal{J} \subset [N+1:N-4B-2U]$,  $\mathcal{K} \subset [N-4B-2U+1:N-2B-2U]$, and $\mathcal{L} \subset [N-2B-2U+1:N-2U]$, $\mathtt{V} = \mathtt{V}_{2L+2H+2M-N}(\bm{b}_{[2L+2H+2M+2B-N]})$, and 
    \begin{align}
        \bm{y} = \begin{bmatrix}
            \Tilde{\bm{W}}_{\theta}(1)\\ \Tilde{\bm{W}}_{\theta}(2)\\ 
            \Tilde{\bm{I}}_2(1)\\ 
            \Tilde{\bm{I}}_2(2)
        \end{bmatrix}.
    \end{align}
    Here, \eqref{comment_2} follows from the data processing inequality, independence between the queries, messages and masking variables, storage security against any $B$ servers and \eqref{helpful_fact_1}.
\end{Proof}

Now, for the second regime, we can write the transmitted symbols as
\begin{align}
   \bm{h}_2=\begin{bmatrix}
        \mathtt{I}_N&0&0\\
        0&\mathtt{V}_{L_1+L_2+\delta+2H+2M+2B-N}(\bm{b}_{[2L+2H+2M+2B-N]})& 0 \\
        0&0& \mathtt{I}_{4B+2U}
    \end{bmatrix}\underbrace{\begin{bmatrix}
        \Tilde{\bm{I}}_1(1)\\
        \Tilde{\bm{I}}_1(2)\\
        \Tilde{\bm{r}}_{[\delta]}\\
        \Tilde{\bm{W}}_{\theta}(1)\\ \Tilde{\bm{W}}_{\theta}(2)\\ \Tilde{\bm{I}}_2(1)\\ 
        \Tilde{\bm{I}}_2(2)\\ 
        \Tilde{\bm{\Delta}}_B(1)\\ 
        \Tilde{\bm{\Delta}}_B(2)\\ \bm{\Delta}_U'(1)\\ 
        \bm{\Delta}_U'(2)
    \end{bmatrix}}_{\bm{x}},
\end{align}
where 
\begin{align}
    &\begin{bmatrix}
         \Tilde{\bm{r}}_{[\delta]}\\ 
         \Tilde{\bm{W}}_{\theta}(1)\\
         \Tilde{\bm{I}}_1(1)\\ 
         \Tilde{\bm{I}}_2(1)\\
         \Tilde{\bm{\Delta}}_B(1) 
    \end{bmatrix}= \begin{bmatrix}
        \bm{r}_{[\delta]}\\
        \bm{W}_{\theta}(1)\\
        \hat{\bm{I}}_{[H+M+B]}(1)\\
        \bm{0}_{2B}
    \end{bmatrix} + \mathtt{CSA}_{N-U, L,N-U}^{-1}\bm{\Delta}_B(1),\\
    & \begin{bmatrix}
          \Tilde{\bm{W}}_{\theta}(2)\\
          \Tilde{\bm{I}}_1(2)\\
          \Tilde{\bm{I}}_2(2)\\
          \Tilde{\bm{\Delta}}_B(2) 
    \end{bmatrix}= \begin{bmatrix}
        \bm{W}_{\theta}(2)\\
        \hat{\bm{I}}_{[H+M+B]}(2)\\
        \bm{0}_{2B}
    \end{bmatrix} + \mathtt{CSA}_{N-U, L,N-U}^{-1}\bm{\Delta}_B(2),
\end{align}
and
\begin{align}
    \hat{\bm{I}}_{[H+M+B]}(i) = \begin{bmatrix}
        \bm{I}_{[H+M]}(i)+\bm{Z}'_{[H+M]}(i)\\
        \bm{R}'_{[B]}(i)
    \end{bmatrix}.
\end{align}
It is clear that the same approach can be used to prove symmetric privacy and eavesdropper protection.

Similarly, for the third regime, the transmitted symbols can be written as
\begin{align}
    \bm{h}_3=\begin{bmatrix}
        \mathtt{I}_N&0&0\\
        0&\mathtt{V}_{L_1+L_2+E}(\bm{b}_{[L_1+L_2+E]})& 0 \\
        0&0& I_{4B+2U}
    \end{bmatrix}\underbrace{\begin{bmatrix}
        \Tilde{\bm{I}}_1(1)\\
        \Tilde{\bm{I}}_1(2)\\
        \Tilde{\bm{r}}_{[E]}\\
        \Tilde{\bm{W}}_{\theta}(1)\\ \Tilde{\bm{W}}_{\theta}(2)\\ \Tilde{\bm{\Delta}}_B(1)\\ \Tilde{\bm{\Delta}}_B(2)\\ \bm{\Delta}_U'(1)\\ 
        \bm{\Delta}_U'(2)
    \end{bmatrix}}_{\bm{x}},
\end{align}
where 
\begin{align}
    &\begin{bmatrix}
         \Tilde{\bm{r}}_{[E]}\\
         \Tilde{\bm{W}}_{\theta}(1)\\
         \Tilde{\bm{I}}_1(1)\\ 
         \Tilde{\bm{I}}_2(1)\\  \Tilde{\bm{\Delta}}_B(1) 
    \end{bmatrix}= \begin{bmatrix}
        \bm{r}_{[E]}\\
        \bm{W}_{\theta}(1)\\
        \hat{\bm{I}}_{[H+T_1+B]}(1)\\
        \bm{0}_{2B}
    \end{bmatrix} + \mathtt{CSA}_{N-U, L,N-U}^{-1}\bm{\Delta}_B(1),\\
    &\begin{bmatrix}
          \Tilde{\bm{W}}_{\theta}(2)\\
          \Tilde{\bm{I}}_1(2)\\
          \Tilde{\bm{I}}_2(2)\\
          \Tilde{\bm{\Delta}}_B(2) 
    \end{bmatrix}= \begin{bmatrix}
        \bm{W}_{\theta}(2)\\
        \hat{\bm{I}}_{[H+T_2+B]}(2)\\
        \bm{0}_{2B}
    \end{bmatrix} + \mathtt{CSA}_{N-U, L,N-U}^{-1}\bm{\Delta}_B(2),
\end{align}
and
\begin{align}
    \hat{\bm{I}}_{[H+T_i+B]}(i) = \begin{bmatrix}
        \bm{I}_{[H+T_i]}(i)+\bm{Z}'_{[H+T_i]}(i)\\
        \bm{R}'_{[B]}(i)
    \end{bmatrix}.
\end{align}

This is also analogous to the previous case with $\delta = E$ and $2H+2M+2B = N$. Thus, the proofs for symmetric privacy and eavesdropper privacy go along the same lines.

Finally, for the last regime, recall that we can write the answers collectively from the $N-U$ serves as follows
\begin{align}
    \begin{bmatrix}
        A_{i_1}^{[\theta]}\\
        \vdots\\
        A_{i_{N-U}}^{[\theta]}
    \end{bmatrix} = &\begin{bmatrix}
        \frac{1}{f_1-\alpha_{i_1}}\!&\!\ldots\!&\!\frac{1}{f_L-\alpha_{i_1}}\!&\!1\!&\!\alpha_{i_1}\!&\!\ldots\!&\!\alpha_{i_1}^{H+M+B-1}\\
        \!\frac{1}{f_1-\alpha_{i_2}}\!&\!\ldots\!&\!\frac{1}{f_L-\alpha_{i_2}}\!&\!1\!&\!\alpha_{i_2}\!&\!\ldots\!&\!\alpha_{i_2}^{H+M+B-1}\\
        \!\vdots\!&\!\ddots\!&\!\vdots\!&\!\vdots\!&\!\vdots\!&\!\ddots\!&\!\vdots\\
        \!\frac{1}{f_1-\alpha_{i_{N-U}}}\!&\!\ldots\!&\!\frac{1}{f_L-\alpha_{i_{N-U}}}\!&\!1\!&\!\alpha_{i_{N-U}}\!&\!\ldots\!&\!\alpha_{i_{N-U}}^{H+M+B-1}
    \end{bmatrix}\bm{x},    
    \end{align}
    where
    \begin{align}
    \bm{x}=\begin{bmatrix}
        \bm{W}_{\theta}\\
        \bm{I}_{[H+M]}+ Z'_{[H+M]}\\
        \bm{R}'_{[B]}\\
        \bm{0}_{2B}
    \end{bmatrix} 
    + \mathtt{CSA}^{-1}_{N-U, L, N-U}\bm{\Delta}_B.
\end{align}
In this regime as well, we can see that, by using the same approach that is used in the first regime, security and privacy requirements are satisfied.

\begin{remark}
    Using the result from \cite{jafar_quantum_unresponsive}, we know that the transfer matrix defined by 
    \begin{align}
        \begin{bmatrix}
            0_N& \mathtt{I}_N
        \end{bmatrix}
            \mathtt{M}^{-1},
    \end{align}
    where $\mathtt{M}$ is as defined in \eqref{main_transfer} or \eqref{answers_collected}, is a valid $N$-sum box transfer matrix. 
\end{remark}

\begin{lemma}\label{lemma9}
    Let $\mathtt{D}_1 = \begin{bmatrix}
        0& \mathtt{I}_N
    \end{bmatrix}\begin{bmatrix}
        \mathtt{G}&\mathtt{H}
    \end{bmatrix}^{-1}$ be a feasible $N$-sum box transfer matrix, then for any $N \times N$ invertible matrices $\mathtt{V}_1$ and $\mathtt{V}_2$, $\mathtt{D}_2 = \begin{bmatrix}
        0& \mathtt{I}_N
    \end{bmatrix}\begin{bmatrix}
        \mathtt{V}_1&0\\
        0&\mathtt{V}_2
    \end{bmatrix}^{-1}\begin{bmatrix}
        \mathtt{G}&\mathtt{H}
    \end{bmatrix}^{-1}$ is a feasible $N$-sum box transfer matrix.
\end{lemma}

\begin{Proof}
    Note that since both $\mathtt{V}_1$ and $\mathtt{V}_2$ are invertible, then $\begin{bmatrix}
        \mathtt{V}_1&0\\
        0&\mathtt{V}_2
    \end{bmatrix}$ is also invertible. Now, since $\mathtt{G}$ is an SSO matrix, we need to check that $\mathtt{GV}_1$ is also an SSO matrix since
    \begin{align}
        \mathtt{D}_2 &= \begin{bmatrix}
        0& \mathtt{I}_N
    \end{bmatrix}\begin{bmatrix}
        \mathtt{V}_1&0\\
        0&\mathtt{V}_2
    \end{bmatrix}^{-1}\begin{bmatrix}
        \mathtt{G}&\mathtt{H}
    \end{bmatrix}^{-1}\\
    &= \begin{bmatrix}
        0& \mathtt{I}_N
    \end{bmatrix}\Big( \begin{bmatrix}
        \mathtt{G}&\mathtt{H}
    \end{bmatrix}\begin{bmatrix}
        \mathtt{V}_1&0\\
        0&\mathtt{V}_2
    \end{bmatrix} \Big)^{-1}\\
    & = \begin{bmatrix}
        0& \mathtt{I}_N
    \end{bmatrix}\Big( \begin{bmatrix}
        \mathtt{GV}_1&\mathtt{HV}_2
    \end{bmatrix}\Big)^{-1}.
    \end{align}
    Thus, $\mathtt{V_1^tG^tJGV_1 = V_1^t(G^tJG)V_1} = 0$ implying that $\mathtt{GV}_1$ is a valid SSO matrix. Now, since $\begin{bmatrix}
    \mathtt{G} & \mathtt{H}
    \end{bmatrix}$ is invertible, thus $\begin{bmatrix}
        \mathtt{GV}_1&\mathtt{HV}_2
    \end{bmatrix} = \begin{bmatrix}
        \mathtt{G}&\mathtt{H}
    \end{bmatrix}\begin{bmatrix}
        \mathtt{V}_1&0\\
        0&\mathtt{V}_2
    \end{bmatrix}$ is also invertible.
\end{Proof}

\bibliographystyle{unsrt}
\bibliography{references.bib}
\end{document}